\documentclass[twocolumn]{aastex61}

\usepackage{multirow}
\usepackage{comment}
\usepackage{amsmath}

\received{Thermidor 1, II}
\revised{Brumaire 18, IV}
\accepted{\today}
\submitjournal{ApJ}

\shorttitle{The IMF in Coma Berenices}
\shortauthors{Gennaro et al.}
\begin{document}

\title{The initial mass function in the Coma Berenices dwarf galaxy from deep near-infrared HST observations.}

\correspondingauthor{Mario Gennaro}
\email{gennaro@stsci.edu}

\author[0000-0002-5581-2896]{Mario Gennaro}
\altaffiliation{Based on observations made with the NASA/ESA {\it Hubble
Space Telescope}, obtained at the Space Telescope Science Institute, which
is operated by the Association of Universities for Research in Astronomy, 
Inc., under NASA contract NAS 5-26555.  These observations are associated
with program GO-13449.}
\affiliation{Space Telescope Science Institute,
3700 San Martin Drive,
Baltimore, MD 21211, USA}

\author{Marla Geha}
\affiliation{Department of Astronomy,
Yale University, New Haven, CT 06520, USA}

\author{Kirill Tchernyshyov}
\affiliation{Department of Physics and Astronomy,
The Johns Hopkins University, 3400 N. Charles Street, Baltimore, MD 21218, USA}

\author{Thomas M. Brown}
\affil{Space Telescope Science Institute,
3700 San Martin Drive,
Baltimore, MD 21211, USA}

\author{Roberto J. Avila}
\affil{Space Telescope Science Institute,
3700 San Martin Drive,
Baltimore, MD 21211, USA}

\author{Charlie Conroy}
\affil{ Harvard-Smithsonian Center for Astrophysics
60 Garden St., MS-19
Cambridge, MA 02138 }

\author{Ricardo R. Mu{\~{n}}oz}
\affil{Departamento de Astronomía, Camino El Observatorio 1515, Las Condes, Santiago, Chile}

\author{Joshua D. Simon}
\affiliation{Observatories of the Carnegie Institution of Washington, 
813 Santa Barbara Street, Pasadena, CA 91101, USA}

\author{Jason Tumlinson}
\affil{Space Telescope Science Institute,
3700 San Martin Drive,
Baltimore, MD 21211, USA}



\begin{abstract}
We use deep $HST$ WFC3/IR imaging to study the Initial Mass Function (IMF) of the ultra faint dwarf galaxy Coma Berenices (Com Ber). Our observations reach the lowest stellar mass ever probed in a resolved galaxy, with 50\% completeness at $\sim 0.17$ M$_{\odot}$. Unresolved background galaxies however limit our purity below $\sim 0.23$ M$_{\odot}$.
If modeled with a single power law, we find that the IMF slope is $-1.45^{+0.29}_{-0.3}$ (68\% credible intervals), compared to a Milky Way value of $-2.3$. For a broken power law, we obtain a low-mass slope of $-1.18_{-0.33}^{+0.49}$, a high-mass slope of $-1.88_{-0.49}^{+0.43}$ and a break mass of $0.57_{-0.08}^{+0.12}$ M$_{\odot}$, compared to $-1.3$, $-2.3$ and 0.5 M$_{\odot}$ for a Kroupa IMF. For a log-normal IMF model we obtain values of $0.33_{-0.16}^{+0.15}$ M$_{\odot}$ for the location parameter and of $0.68_{-0.12}^{+0.17}$ for $\sigma$ (0.22 M$_{\odot}$ and 0.57 for the Chabrier system IMF). All three parametrizations produce similar agreement with the data. Our results agree with previous analysis of shallower optical \textit{HST} data. However analysis of similar optical data of other dwarfs finds IMFs significantly more bottom-light than in the Milky Way. These results suggest two, non mutually exclusive, possibilities: that the discrepancy of the dwarf galaxies IMF with respect to the Milky Way is, at least partly, an artifact of using a single power law model, and that there is real variance in the IMF at low masses between the currently studied nearby dwarfs, with Com Ber being similar to the Milky Way, but other dwarfs differing significantly.
\end{abstract}



\section{Introduction} 
\label{sec:intro}

In recent years the question of the universality of the initial mass function (IMF) across different environments as well as throughout cosmic history has seen the rise of a new interest, thanks to several new results. \cite{2010Natur.468..940V,2011ApJ...735L..13V,2012ApJ...760...70V} and \cite{2012ApJ...760...71C} have used integrated spectra of giant elliptical galaxies to claim that these environments show a more bottom heavy IMF with respect to the Galaxy \citep[see also][]{2012ApJ...753L..32S}.
\cite{2012Natur.484..485C} claim a similar bottom-heaviness for the IMF of early type galaxies using a different approach based on the comparison between masses derived by dynamical models (constrained by integral-field spectroscopy) and masses derived from stellar population synthesis (constrained by photometric measurements).
More recently, these results have been questioned by \cite{2014MNRAS.443L..69S} who compare the spectroscopic and dynamical methods.
%They compare compare the results of the spectroscopic method of \cite{2010Natur.468..940V,2011ApJ...735L..13V,2012ApJ...760...70V} and \cite{2012ApJ...760...71C} and the dynamics-based one by  \cite{2012Natur.484..485C} using galaxies common to both samples.
While the global trends from the two approaches are similar, such detailed comparison shows large systematics and lack of agreement for the objects that can be studied using both approaches, thus questioning the reliability of the individual results.
Furthermore, \cite{2017ApJ...845..157N} compare dynamical, spectroscopic, and, in addition, lensing-based IMF estimates for a sample of three nearby giant ellipticals. 
Their results are that when the underlying IMF is parametrized as a single or broken power law, the results for the three methods disagree for 2 out of the 3 galaxies in the sample.
However, using more flexible, including non-parameteric, IMF forms partly relieves this tension.
It must be noted however that the mismatch pointed out by \cite{2014MNRAS.443L..69S} might be due to the existence of radial gradients in the IMF of giant ellipticals \citep{2015MNRAS.447.1033M,2016MNRAS.463.3220L,
2017ApJ...841...68V}. Thus the differences in individual galaxies results can be related to the use of different apertures between the different sets of observations.

While studies of integrated light from external galaxies allow us to probe a large range of environments, studies based on resolved star counts can help reaching a finer level of detail, less dependent on modeling assumptions.
On the other end, the majority of resolved studies of the IMF at low stellar masses have been limited to the nearby
Galactic field and star clusters, which do not reflect the wide range of environments over which the IMF is routinely applied. 

For the Milky Way field, \cite{1955ApJ...121..161S} originally proposed a single
power-law IMF, but modern observations show a break in the IMF slope at low masses
\citep[for a review, see][]{2010ARA&A..48..339B}. The IMF is parameterized as either a log-normal
distribution with a turn-over mass near 0.25M$_{\odot}$ \citep{2003PASP..115..763C,2010AJ....139.2679B}, or
a broken power law, $\mathrm{d}N/\mathrm{d}m \propto m^{\alpha}$, with slope ${\alpha} = −2.3$ above 0.5~M$_{\odot}$, and shallower slope of ${\alpha} = -1.3$ below \citep{2002Sci...295...82K}. The stellar mass at which this slope transition occurs is predicted to depend on the physical properties of the stellar birth cloud \citep{2005MNRAS.359..211L}.

Milky Way open clusters provide a relatively large
stellar mass range over which to measure the IMF \citep[0.08 to 100, the upper limit depending on the cluster's age M$_{\odot}$, see e.g.][]{2003A&A...400..891M,2012ApJ...748...14D,2012A&A...540A..57H,2013ApJ...762..123W,2017A&A...602A..22A},
but are predominantly metal-rich and only the youngest are not affected by mass segregation and evaporation. \textit{HST} studies of Galactic globular clusters probe the IMF
down to main sequence masses between 0.1 and 0.8~M$_{\odot}$ \citep{2010AJ....139..476P,2012MNRAS.422.1592L,2017MNRAS.471.3668S,2017MNRAS.471.3845W}, %suggesting an IMF similar to the Milky Way field itself. 
and find slope values covering a wide range from -1.5 to +0.2, thus encompassing the mean value estimated in the Galactic field.
However, dynamical evolution, such as mass segregation
and evaporation, preferentially affects low-mass stars and can significantly change the slope
of the mass function in these concentrated systems \citep{1997MNRAS.289..898V}. Dwarf galaxies
provide a similarly low metallicity environment, but have relaxation times many times the
age of the Universe, therefore minimizing any corrections for dynamical evolution \citep[see Section 6.1 of][]{2018ApJ...855...20G}.

Several recent studies exist that try to address the question of the IMF universality (or lack thereof) in nearby dwarf Milky Way satellite galaxies. %These studies are based on direct counts of individual, resolved stars, and while still subject to a series of difficulties, are less prone to systematics due to the modeling of integrated properties. 
Given their different properties (e.g., morphology, metallicity, star formation history), these dwarf satellites are interesting targets to explore possible IMF differences with respect to the Milky Way.

\begin{comment}
These studies usually adopt a single power law model for the IMF, parametrized by a slope, and compare their best fit results to an average value for the Milky Way, typically $-2.3$ -- $-2.35$ \citep{1955ApJ...121..161S,2001MNRAS.322..231K}.
The IMF in our Galaxy is observed to experience a flattening, or turnover, below $\sim0.5$M$_{\odot}$ \citep{2001MNRAS.322..231K,2003PASP..115..763C}, so results of these studies need to be taken with great care when attempting a comparison with the Milky Way. The single power law parametrization is however convenient, and allows ease of comparison with, e.g., the extragalactic studies and ee will adopt it throughout this paper, together with a Log-normal parameterization.

\end{comment}

One of the first of such studies is the one by \cite{2002NewA....7..395W}, who measured an IMF slope of $-1.8$ for the Ursa Minor dwarf spheroidal in the 0.4 -- 0.7 M$_{\odot}$ range, using $HST$ WFPC2 data.
Recently \cite{2013ApJ...763..110K}, used $HST$ ACS/WFC to reach $\sim$0.4~M$_{\odot}$ in the Small Magellanic Cloud outskirts. They found that the IMF there has a slope of $\sim -1.9$, shallower than the typical Salpeter value of $-2.35$.
\cite{2013ApJ...771...29G} also used ACS/WFC to study the IMF of the Hercules and Leo~IV ultra-faint dwarf galaxies (UFDs), measuring slopes of $-1.2$ and $-1.3$ respectively.
\cite{2018ApJ...855...20G} extended the analysis of \cite{2013ApJ...771...29G} adding 4 more UFDs to the sample, namely Bo\"{o}tes~I (Boo~I), Canes~Venatici~II (CVn~II), Coma~Berenices (Com~Ber) and Ursa~Major~I (UMa~I). They found similar results to \cite{2013ApJ...771...29G} for Hercules and Leo~IV, and, when using a single power law model, they found an overall shallower slope for the UFDs IMF when compared to the Salpeter value. 
\cite{2017MNRAS.468..319E} note that a flatter than Salpeter slope could be a natural result if the true, underlying IMF is in fact a log-normal. A single power law fit to masses drawn from a \cite{2003PASP..115..763C} IMF, limited to a mass range of 0.4-0.8 M$_\odot$ \citep[the one accessible in the][studies]{2013ApJ...763..110K,2013ApJ...771...29G,2018ApJ...855...20G}, gives in fact a slope $\alpha=-1.55$.

Within their sample, \cite{2018ApJ...855...20G}  noted that for the 2 closest UFDs, Boo~I and Com~Ber, a Salpeter value could be ruled out only at 1$\sigma$ level, while for the other UFDs the discrepancy was larger, up to more than 3$\sigma$ for Hercules and Leo~IV.
Moreover, when modeling the IMF with a log-normal function, the results for the overall sample were more consistent with the Milky Way values \citep{2003PASP..115..763C,2010AJ....139.2679B}, with again Boo~I and Com~Ber showing the least dissimilarity with respect to the Galaxy.

The typical limiting mass of the \cite{2018ApJ...855...20G} ACS data was 0.4 -- 0.5 M$_{\odot}$. 
In the current work we present a study of the IMF of Com~Ber, the closest of the 6 UFDs in the ACS sample, using deeper infrared data. 
ComBer is a very faint system, with $M_V \sim -3.8$ mag. Its mean metallicity is [Fe/H] $=-2.6$ dex, with a spread of about 1 dex \citep{2014ApJ...796...91B}. From optical data its distance has been determined to be $\sim 43$ kpc \citep{2014ApJ...796...91B}. Its half-light radius is 77 pc, which at its distance subtends an angle of 6.15 arcminutes.

We have targeted Com~Ber as part of an $HST$ program (GO:13449, PI: M. Geha) which used the infrared channel of the WFC3 camera, better suited for efficient observations of faint, cool stars. Using WFC3, our observations reached down to $\sim 0.2$M$_{\odot}$ in Com~Ber, thus allowing the deepest extra-Galactic study of the IMF so far.
Com~Ber shows neither kinematic nor photometric evidence for tidal disruption \citep{2007ApJ...670..313S, 2010AJ....140..138M}. While its proximity to the Milky Way makes it vulnerable to tides, these processes should not bias the IMF, since  
Com~Ber 2-body relaxation time is of the order of $5\times 10^3$ Gyr \citep[see equation (5) of][]{2018ApJ...855...20G}, and thus it is not expected to suffer from mass-dependent losses. 
%The Com~Ber relaxation time is much longer than the age of the Universe, therefore the velocity distribution of Com~Ber stars does not depend on their mass, hence neither does their escape or stripping probability. 
Of the nearest UFDs, Com~Ber has the highest surface brightness ($\mu_V$ = 27.3 mag arcsec$^{−2}$), and thus is a very suitable candidate for a deep IMF study.

\section{The data}
The data used in this paper were obtained as part of the HST GO program 13449 (PI: M. Geha). A total of 44 WFC3/IR orbits were allocated to this program, with an equal number of ACS/WFC parallel images; the latter target the outskirts of Com~Ber. This work only deals with the WFC3/IR images, covering the center of Com~Ber.

\subsection{Observing strategy}
We utilized the WFC3/IR instrument in the F110W and F160W filters.
We imaged the galaxy in a 2x2 tiles mosaic, and reduced and analyzed the 4 tiles independently of each other.
The program consisted of 44 \textit{HST} orbits, equally divided among the 4 tiles.
The 11 orbits per tile were split with 4 orbits in F110W and 7 in F160W.
Each orbit was split in 2 exposures, thus resulting in 8 dithered exposures in F110W and 14 dithered exposures in F160W per tile.
The resulting exposure times are 10994 seconds in F110W and
18989 seconds in F160W per tile.

\subsection{Data reduction}

WFC3/IR images in filters that are sensitive to the Earth's atmosphere He\textsc{i} 1.083 $\mu$m line afterglow
can suffer from a time-variable background.
This is due to the change in the depth of the atmosphere along the line of sight and is especially severe at ingress/egress of $HST$ from occultation during its orbit around the Earth.
This problem may affect F110W observations, and thus
impact the up-the-ramp fitting performed by the \texttt{calwfc3} pipeline, which assumes a constant background, thus producing
images with artificially increased noise. 
To reduce the problem, the individual up-the-ramp non-destructive reads of a WFC3/IR \textsc{Multiaccum} exposure can be equalized by subtracting the average sky value measured from the difference between two consecutive reads.
We used software developed by a former WFC3 team member (B. Hilbert, \emph{priv. comm.})
to treat the images before running the standard \texttt{calwfc3} reduction process \citep[see also][for a summary on the He~\textsc{I} related problem and a very similar mitigation strategy, as well as publicly available software performing similar pre-processing of WFC3/IR images]{2016wfc..rept...16B}.
%The WFC3 team has developed a correction to this problem thatlooks for non-linear behavior in the up-the-ramp fitting and removes any extra signal. 
%This correction is applied to raw images before they are run through the standard CALWF3
%pipeline. The \texttt{wfc3\_remove\_variable\_background.pro} provided by the team (Version 1.0,
%6 May 2014) was used.

%We also used an updated bad pixel table on these images, including flags for new blobs that have appeared on the detector. 

The images were aligned and drizzled using the standard Drizzlepac tools \citep{2012drzp.book.....G}, using the F110W
band as the astrometric reference. Images were drizzled to rotation=0 (North up), a scale
of 60~mas/pix, and pixfrac of 0.8. The latter controls the fraction by which the input pixels are “shrunk” before being drizzled onto the output image grid, i.e. the drizzle "drop" size. The final drizzled images were converted to counts images
in preparation for use with the stand-alone \texttt{DAOPHOT-II} package \citep{1987PASP...99..191S}.

\subsection{Photometry}
\label{sec:phot}

\begin{figure}
\includegraphics[width=0.5\textwidth]{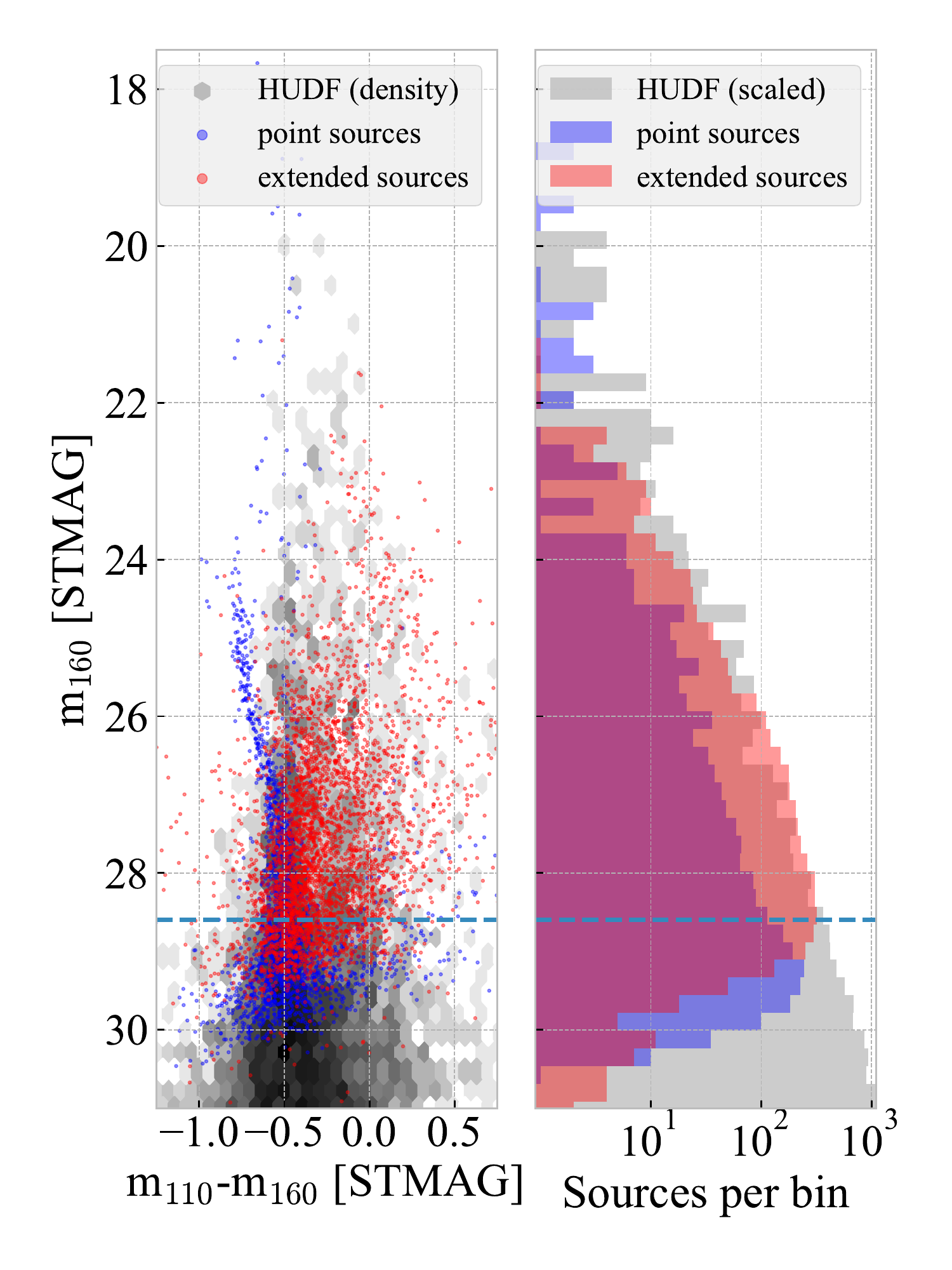}
\caption{Left: Color-magnitude diagram showing sources identified as point sources (blue) and as extended ones (red), with superimposed a density plot of the Hubble Ultra deep field objects (grey scale, see text for more details). Right: m$_{160}$ luminosity function for the same objects. The dashed blue line at m$_{160}$ = 28.6 mag indicates where our background objects counts depart from the HUDF counts; this is where we set our limiting depth for further analysis. In the right panel the HUDFs counts have been normalized to the area of our mosaic. \label{fig:contam}}
\end{figure}

PSF fitting photometry was conducted using the stand-alone \texttt{DAOPHOT-II} package, to obtain F110W and F160W magnitudes calibrated in the STMAG system.
Thresholds in the $\chi$ and sharpness values are used to distinguish point sources from resolved background objects.
The combined color-magnitude diagram (CMD) for the 4 tiles is shown in Figure~\ref{fig:contam}. 

To test our ability to resolve background sources, we compare the luminosity function of resolved objects, i.e. all the photometric detections that do not make the $\chi$ and sharpness cut, to that of the Hubble Ultra deep Field (HUDF), using
the catalog made available by \cite{2015AJ....150...31R}. Because there is no F110W observation for the HUDF, we mock F110W data by linearly interpolating the fluxes in the F105W and F125W.

Figure~\ref{fig:contam} shows the HUFD sources, scaled to the total area of our observations. The luminosity function of our extended, background sources matches that in the HUDF very well down to about F160W = 28.6 mag, below which our extended background source luminosity function falls below the HUFD one. This means that we fail to characterize background objects as extended sources beyond this point. Even though our 50\% completeness limit for point sources is about 0.8 mag fainter than 28.6 mag in F160W (see Section~\ref{sec:AStests}), we choose to preserve the purity of our catalog by limiting our analysis to stars brighter than F160W = 28.6 mag, where we can safely rely on our $\chi$-sharpness selection.
This magnitude value corresponds to about 0.23 M$_\odot$ at the distance and extinction of Com Ber. The mass at the 50\% completeness limit, F160W $\sim$ 29.4 mag, is instead about 0.17 M$_\odot$, thus the loss in mass range corresponding to the adopted conservative cut in magnitude is small but not negligible.
%In a standard $\Lambda$CDM cosmological model, a structure of 1 kpc subtends an angle of 0.12'' at $z\sim1.5$ (the redshift at which the angular distance reaches a minimum), comparable to the size of a WFC3/IR pixel (0.13''), and the size of $\theta \sim \lambda/D$ at 1.5 $\mu m$, the F160W filter effective wavelength (also 0.13''). 
The half-mass radius for a typical dwarf galaxy, 300 pc, is unresolved in WFC3/IR beyond $\sim1$ Gpc, corresponding to a distance modulus of 40 mag and to a redshift, $z\sim 0.15$. Dwarf galaxies of intrinsic brightness $M_{160} \sim -10$~mag at such a distance contribute to the unresolved background sources seen in Figure~\ref{fig:contam}, with brighter dwarfs galaxies contributing up to even larger distances. 
This problem will be alleviated (by a factor of 3 at fixed wavelength) when the \textit{James Webb Space Telescope} will become operational.

We further select our data by excluding the sources on the red giant branch (RGB). In practice, we adopt a bright magnitude cut at F160W = 23 mag.
This cut allows us to ignore the RGB  region, where the stellar models tend to not be able to reproduce the observed colors, in our simulations versus data comparison. Given the faster evolution on the RGB relative to the main sequence,
this bright limit has no significant impact on the mass range probed.
The final number of selected stars per tile is 223, 163, 149, and 182 respectively, for a total of 717.

\begin{figure}
\includegraphics[width=0.5\textwidth]{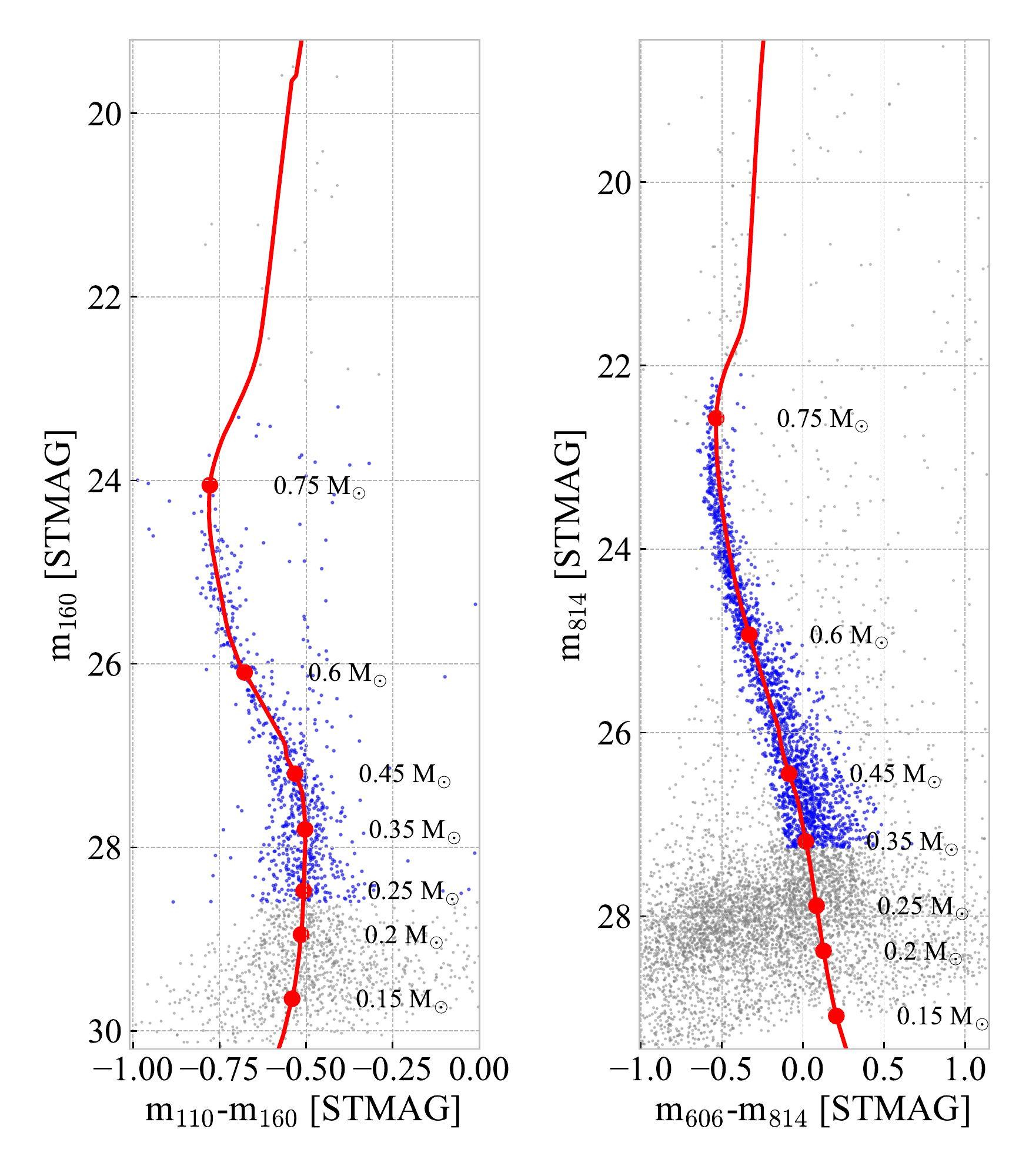}
\caption{Left: IR data used in this paper. Right: optical data from \cite{2014ApJ...796...91B}, used for deriving the IMF in \cite{2018ApJ...855...20G}. The grey points indicate all the objects identified as point sources. The blue points represent the data used for inferring the IMF in either this work or \cite{2018ApJ...855...20G}. In red is a 14.0 Gyr, [Fe/H] = -2.6 dex isochrone from \cite{2014ApJ...794...72V}, transformed into the appropriate passbands using the MARCS model atmospheres \citep{2008A&A...486..951G}. For the IR isochrone we apply the corrrection described in Section~\ref{sec:models}. The red points mark several mass values along the isochrone. The y-axes in the two panels have the same total magnitude range and are arranged vertically to get the same y-position for the 0.45 M$_{\odot}$ model.\label{fig:CMDOPTIR}}
\end{figure}

The selected data are highlighted in a CMD in Fig.~\ref{fig:CMDOPTIR}, where the IR data presented in this paper are also compared to the optical Com~Ber data first presented by \cite{2014ApJ...796...91B}, and used for deriving the IMF in \cite{2018ApJ...855...20G}. The figure shows how the new data reach a lower stellar mass, thus giving more mass leverage for the IMF determination with respect to the existing optical data. The optical observations cover however a larger area on the sky, and thus the number of Com~Ber members is larger in the optical than in the IR CMD.

%\begin{deluxetable}{c|c}
%\tablecaption{Number of stars in each tile after the selection described in Sect.~\ref{sec:phot}.\label{tab:nstars}}
%\tablehead{ \colhead{Tile} & \colhead{Number of stars} }
%\startdata
%1 & 223 \\
%2 & 163 \\
%3 & 149 \\
%4 & 182 \\
%\hline
%Total & 717 \\
%\enddata
%\end{deluxetable}

\subsection{Artificial star tests}
\label{sec:AStests}
We perform artificial star tests to obtain an estimate of the noise distribution of our sources. We use these experiments in our synthetic color-magnitude diagram simulation when fitting for the IMF parameters.
The artificial star tests were produced by randomly populating isochrones at [Fe/H]=$-1.6$,
$-2.0$, and $-2.4$ dex and age = 10, 11, 12, and 13 Gyr. Each pass of artificial star tests inserted 1000
stars into the images and blindly recovered them, comparing output to input. Specifically,
each pass drew 800 stars from the isochrones, with significant scattering (to fill in the CMD
parameter space), plus another 200 stars at random locations over the entire CMD (to avoid
having gaps in the artificial star CMD coverage). The result places the most weight where
we need it (i.e., near real isochrones, where most stars should fall) but characterizes the
uncertainties and incompleteness over a broad swath of CMD space. The same $\chi$ and sharpness criteria applied to the real stars are applied to the artificial stars, and if the latter do not pass the $\chi$-sharpness test, they are considered undetected.

\begin{figure}
\includegraphics[width=0.5\textwidth]{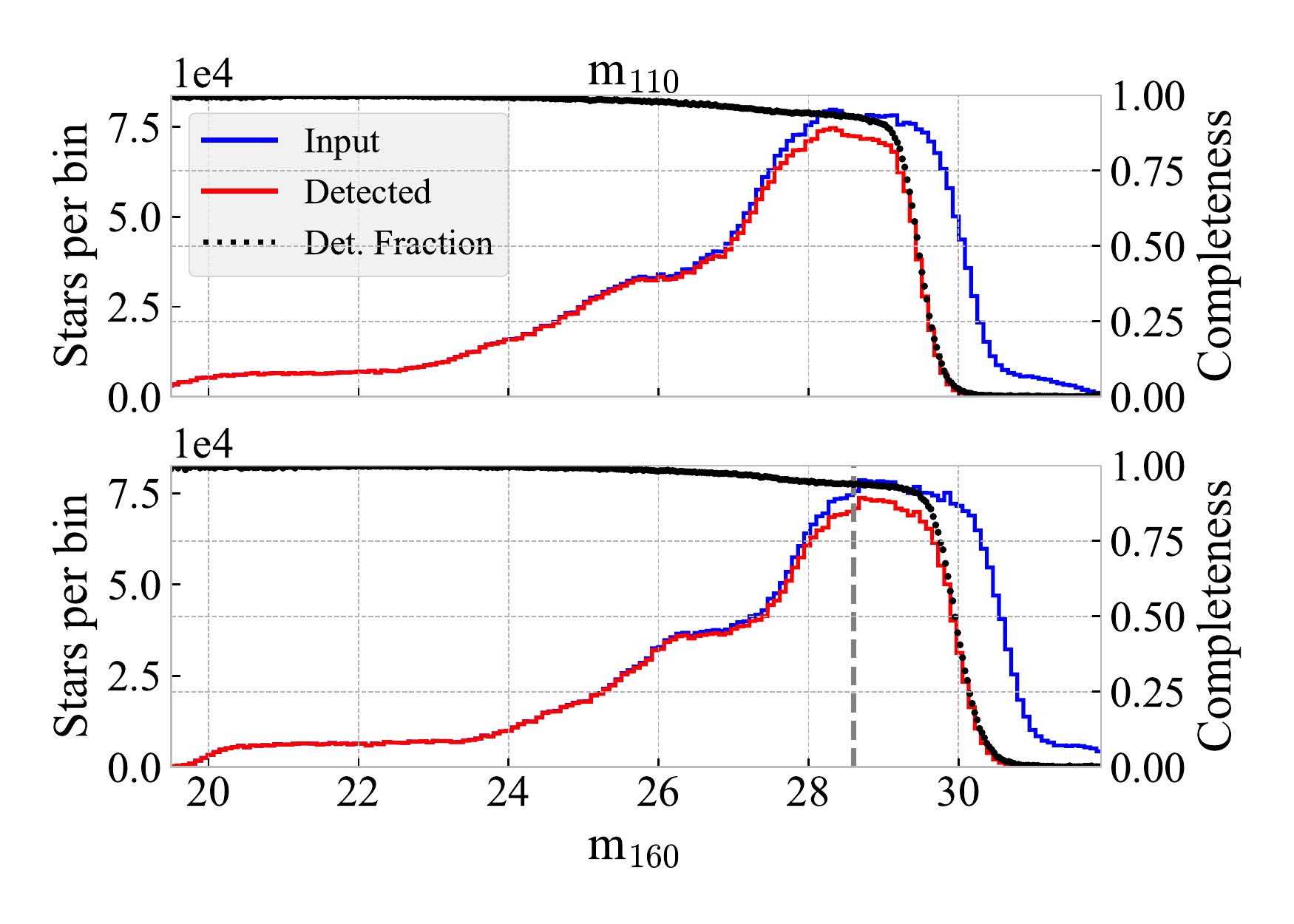}
\caption{Luminosity distributions of the input magnitudes of the artificial stars (blue), and the same but only for stars that have a valid output magnitude (red); all 4 tiles are shown together. Superimposed, black dotted line, on a different y-axis is the ratio between the 2, i.e. the completeness fraction as function of the input magnitude of the artificial stars. The vertical dashed line at 28.6 magnitudes, in the  $m_{160}$ panel (bottom), shows the adopted magnitude limit for which extended sources can still be robustly identified, and which is well above the 50\% completeness limit. 
\label{fig:ASTests}}
\end{figure}

Figure~\ref{fig:ASTests} shows the luminosity functions of the inserted stars (input magnitudes in each band) and the luminosity function of the detected ones (again input magnitudes), as well as the ratio of the two. The vertical dashed line in the $m_{160}$ bottom panel corresponds to 28.6 magnitudes, i.e., the limit at which we loose the ability to detect extended background sources in our data. This is the faint cut we have to impose to our catalog, in order to avoid contamination from unresolved background galaxies. It is clear that our ability to detect point sources extends well past this limit, with the 50\% completeness level being about 29.4 magnitudes in F160W. However we choose to keep the highest possible purity in our catalog, at the cost of loosing some depth.

\subsection{Residual Photometric Uncertainty}
\label{sec:extraerror}
During the analysis stage we observed that the photometric uncertainty encoded in the artificial stars was slightly underestimated.
In particular, the scatter of simulated color magnitude diagrams, built using the artificial star tests and realistic metallicity and age distributions for ComBer, showed a narrower main sequence than that observed.
This extra error may be due to the limitations of the PSF model constructed from the undersampled data.
An other possibility is a metallicity effect. While we are using the most up-to-date metallicity distribution function (MDF) available for Com Ber \citep{2014ApJ...796...91B}, if the true intrinsic distribution is slightly broader, it may produce extra CMD width, with respect to what we can reproduce using our adopted MDF.

We determined that an extra error term of 0.0175 mag (standard deviation) in each photometric band was enough to force the simulations to agree with the data, in terms of simulated main sequence width.
This extra error is implemented in our subsequent analysis by further scattering each artificial star output magnitude using a draw from a normal distribution with width equal to 0.0175 mag.
It is worth noting that this error term is quite small, on the order of the photometric repeatability quoted by the WFC3/IR 
instrument team ($\sim$0.5-1\%, see the WFC3 data handbook\footnote{\cite{2018wfc3.book.....R}, available at \url{http://www.stsci.edu/hst/wfc3/documents/handbooks/currentDHB/wfc3_cover.html}}), and thus our heuristic approach of adding an extra random Gaussian term is appropriate and sufficient for our purposes.

\section{The models}
\label{sec:models}
We adopt the same $\alpha$-enhanced stellar models as in \cite{2014ApJ...796...91B} computed using the 
Victoria-Regina evolutionary code \citep[see][and references therein]{2014ApJ...794...72V}, in the metallicity range $-4.0 < $~[Fe/H]~$ <-1.0$ dex. These models have [$\alpha$/Fe] = +0.4 dex, and a further oxygen enhancement, $\Delta$[O/Fe], that increases at decreasing [Fe/H] values \citep[details in Section 3.2 of][]{2014ApJ...796...91B}.
We transform the models into the observational plane using tables of synthetic fluxes computed with the MARCS 
code \citep{2008A&A...486..951G}, and the most up-to-date WFC3 filter throughputs.
The \cite{2014ApJ...796...91B} oxygen-enhanced models were computed ad-hoc and extend only down to 0.4~M$_\odot$.
In order to further extend the grid of models to lower masses, we computed $\alpha$-enhanced models without extra $\Delta$[O/Fe]. These models are then "patched" by forcing them to agree with the oxygen-enhanced ones at 0.4~M$_\odot$, and by assuming that a rigid translation applies to the model for all masses below 0.4~M$_\odot$. 

\begin{figure*}
\includegraphics[width=.985\textwidth]{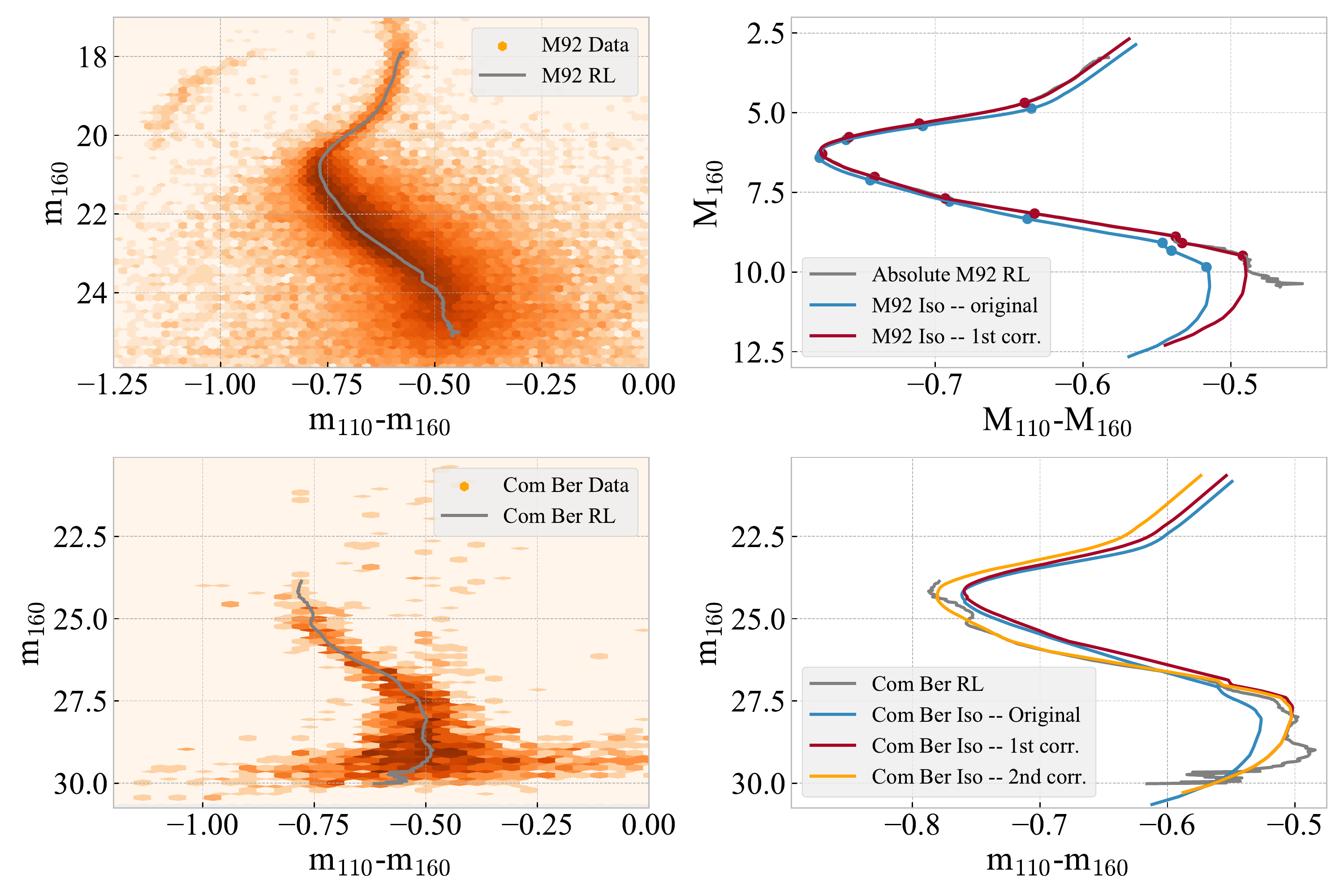}
\caption{Calibration of the isocrone library with respect to the M92 ridge line (top panels) and the Com Ber ridge line (bottom panels). The density plots on the left column are on a logarithmic scale.
% * <tbrown@stsci.edu> 2017-11-07T18:12:43.392Z:
% 
% Why "bulk" ?
%MG: right, why bulk? I shall remove that
% 
% ^.
\label{fig:isocal}}
\end{figure*}

\subsection{First empirical correction: M92 ridge line}
\label{sec:fec}
\cite{2014ApJ...796...91B} showed a very good agreement of the stellar models with the M92 cluster ridge line in the HST/ACS optical (F606W-F814W, F814W) color-magnitude diagram. M92 is the most metal-poor cluster for which we have good WFC3/IR data from the HST Galactic Bulge Treasury Program (GO: 11664, PI: T.~M.~Brown). Such data can be used for calibration of the stellar models in the F110W and F160W bands.
We do not observe the same agreement of the models with the M92 IR data. However, rather than fine-tuning the isochrone parameters to try and obtain a better fit of the M92 ridge line, we assume that the cluster parameters for which the models give good agreement in the optical are valid: $(m-M)_{0} = 14.62$~mag, E(B-V)= 0.023 mag, age = 13.2 Gyr, [Fe/H] = -2.3 dex.
We derive an empirical calibration of the IR models with respect to the M92 ridge line in the following steps:
\begin{enumerate}
\item we generate a ridge line for the  M92 (F110W,F160W) data, using the technique described in \cite{2016ApJ...823...18C}, this step is shown in Figure~\ref{fig:isocal}, top right
\item we subtract the extinction in each band and distance modulus from the M92 ridge line using $(m-M)_{0} = 14.62$~mag, E(B-V)= 0.023 mag, $R_V$=3.1 and assuming that the extinction law can be described by a \cite{1989ApJ...345..245C} model. This results in $(A_{110},A_{160}) = (0.0228, 0.0144)$ mag at the respective effective wavelengths of each of the F110W and F160W filters. The distance and extinction subtracted ridge line is shown as "Absolute M92 RL" in the top-right panel of Fig.~\ref{fig:isocal} 
\item We compute the isocrone for the nominal M92 values of age (13.2 Gyr), [Fe/H] (-2.3 dex) and [$\alpha$/Fe] (0.4 dex), with [O/Fe] = +0.64 dex. This is shown as "original" in Figure~\ref{fig:isocal}, top right
\item we identify "characteristic points" on both the M92 ridge line and the original isochrone. These are visible features in the 2 curves, and we assume that, for a calibrated model, their color and magnitudes need to match exactly. These characteristic points are marked by dots in the top right panel of Fig.~\ref{fig:isocal} (the dots for the corrected model lie, by definition, on the ridge line). Some of these points are identifiable with significant evolutionary stages (i.e., the main sequence turn-off, the low-mass main sequence kink, the base of the red giant branch). Others have been chosen based on morphological characteristics of the isochrones (e.g the mid-point in the sub-giant branch, slight bends along the main sequence between the kink and turn-off)   
\item we compute the differences in F160W magnitude and F110W-F160W color between the points corresponding to the same stages along two curves 
\item we interpolate such correction as function of $\log g$ in the model
\item assume that the same correction in $\log g$ applies for all ages and metallicities 

\end{enumerate}

\subsection{Second empirical correction: Com Ber ridge line}
\label{sec:sec}
The above sequence of steps does not yet provide a complete agreement when the models for the peak of the Com Ber metallicity distribution function are computed and overplotted on the data. 

We thus define a second calibration step, this time in apparent magnitude space. The steps are the following:
\begin{enumerate}
\item build a ridge line for the Com~Ber population. This is shown in the bottom left panel of Figure~\ref{fig:isocal}
\item compute the isochrone for the bulk of the Com Ber population, corresponding to the peak of the metallicity distribution of the galaxy, [Fe/H] $= -2.6$ dex and at an age of 14.0 Gyr, and at its distance, (m-M)$_0$= 17.956 mag, and extinction, $A_\mathrm{V} = 0.124$ \citep[all values from][]{2014ApJ...796...91B}. From $A_\mathrm{V}$ we derive $(A_{110},A_{160}) = (0.0398,0.0251)$ mag.
This is the isochrone labeled as "original" in the lower-right panel of Fig.~\ref{fig:isocal}
\item apply the empirical calibration described in Sect.~\ref{sec:fec} to the original isochrone thus obtaining the isochrone labeled as "1st corr." in the lower-right panel of Fig.~\ref{fig:isocal}
\item compute residual color offsets between the "1st corr." isochrone and the Com~Ber Ridge Line 
\item interpolate the residual offset as function of m$_{160}$ 
\end{enumerate}

The combination of the two calibration steps gives us models that agree very well with the Com~Ber ridge line. The residual offsets visible in the lower right panel of Figure~\ref{fig:isocal}, between the final model (orange) and the ridge line(gray) are of the order of $\lesssim 0.01$ mag.

It must be noted that the extent of the discrepancy between the uncorrected models and the M92 or Com~Ber ridge lines is of the order of 0.05 mag maximum, and only in the coolest regions of the CMD. This is probably related to uncertainties in the model atmospheres of cool stars in the near-IR.  Around 0.3 M$_{\odot}$, where the most serious discrepancies arise, the mass-luminosity relation in our passbands is of about 0.1 M$_{\odot}$/mag, thus an error of 0.05 mag in color, would correspond to only 5 thousandths M$_{\odot}$.
The impact of such errors on the IMF determination is negligible.

\section{The method}

Our analysis aims at deriving the best parameters for an underlying model describing the IMF of Com~Ber. We assume three possible IMF parametrizations: a single power law (SPL), a broken power law (BPL), and a log-normal function (LN).
The models functional forms are thus specified:

The mass distributions are specified as
\begin{align}
p_{\mathrm{SPL}}(m|\alpha)  \propto & \; m^{\alpha} \\
p_{\mathrm{BPL}}(m|\alpha_1,\alpha_2, m_{brk})  \propto & \;m^{\alpha_1}, m> m_{brk} \tag{2a}\\
& \; km^{\alpha_2}, m<m_{brk} \tag{2b}\\
& \;  k = m_{brk}^{\alpha_1-\alpha_2}\nonumber\\
p_{\mathrm{LN}}(m|m_c, \sigma)  \propto &  \; \frac{1}{m} e^{-\frac{1}{2}\left(\frac{\log m-\log(m_c)}{\sigma}\right)^2 \label{eq:ln}} \tag{3}
\end{align}

The method we use to fit for the IMF parameters is presented in \cite{2018ApJ...855...20G} and it is based on a combination of Approximate Bayesian Computation (ABC) techniques, Kernel Distance methods and Markov Chain Monte Carlo sampling.
The statistical model used to describe the observed CMD is defined in the framework of Poisson Point Processes \cite[see][and references therein]{2015ApJ...808...45G}.

In brief, the method consists of i) moving in the parameter space by drawing sets of parameters, using existing Markov Chain Monte Carlo methods (we adopt the \citeauthor{2013PASP..125..306F}, \citeyear{2013PASP..125..306F} implementation of the \citeauthor{2010CAMCS...5...65G}, \citeyear{2010CAMCS...5...65G} affine-invariant sampler). As an example, for the single power-law model, the parameters set would consist of a total number of stars, a slope, and a binary fraction, ii) building a synthetic CMD based on those parameters, and using the stellar models described in Sect.~\ref{sec:models} and the artificial stars experiments described in Sect.~\ref{sec:AStests}, iii) comparing the simulated and observed data sets by defining a suitable distance between them.
The distance measurement is based on kernel distance methods \citep[see][for more details]{2018ApJ...855...20G}, iv) define a maximum acceptable distance threshold, retaining parameter sets for which the distance is below the threshold. The fact that the threshold is small, but not zero, makes this method an Approximate, and not an exact Bayesian Computation.

The ensemble of accepted parameter sets is an approximation of their posterior probability density function.
Each parameter set specification includes the specification of the total intensity, i.e., the integral of the IMF in the mass interval under consideration. The boundaries of the integral, in our case are $[0.175,8] \mathrm{~M}_{\odot}$. The intensity can be regarded as a normalization factor.
The total number of generated stars is a drawn from a Poisson distribution with mean equal to the total intensity.

Similarly to \cite{2018ApJ...855...20G} we specify the IMF in terms of the system mass. For single stars, this is equivalent to the stellar mass. In the case of binaries, the system mass is equal to the sum of the primary and secondary mass. We assume that the binary mass ratio (i.e., the ratio of the secondary to the primary mass) is uniformly distributed between 0 and 1. While the mass ratio distribution is fixed, the binary fraction is a parameter of our IMF models.

In building the synthetic CMDs that are compared to the observed ones, we adopt the star formation history and metallicity distribution function derived for Com~Ber in \cite{2014ApJ...796...91B}. We also adopt their distance modulus and E(B-V) reddening. The latter is converted into F110W and F160W extinction values using the \cite{1989ApJ...345..245C} extinction law (see Section~\ref{sec:sec} for the numerical values of these parameters).

\begin{comment}
In Paper I we have illustrated how the choice of the prior on the total number of stars can in some cases affect the results on the IMF parameters. 
In that work we were reaching only down to 0.5 M$_{\odot}$, and very shallow IMF slopes were allowed by the lack of constraints in the very low mass regime. Shallow slopes resulted in very large intensities (or large number of stars being born), accounting for the fact that the many stars born above the turn-off mass are no longer observable. A logarithmic prior on the intensity was helping suppress the very large intensities, thus the shallow slopes tail of the distribution. At the same time, the logarithmic prior choice was not just convenient, but also a well justified choice: it is an observed fact and a prediction of cosmology that the Universe contains more small galaxies than large ones, thus, a priori, a lower number of stars is expected.
In the present work, the likelihood is better constrained due to the deeper observations, and using a uniform prior in N or in $\log$ N makes a very small difference in the resulting IMF slope. 
We will thus only report results for the logarithmic prior, which seems a better justified choice.
\end{comment}

\begin{deluxetable*}{l|c|cccc}
\tablecaption{Best-fit parameters and credible intervals for the 3 IMF models. The credible intervals are defined as containing $(0.6827, 0.9545, 0.9973)$ of the posterior.\label{tab:res}}
\tablehead{
\colhead{Model} & \colhead{Parameter} & 
\colhead{Average} &\colhead{68\% Cr.I.}       & \colhead{95\% Cr.I.} & \colhead{99\% Cr.I.} }
%\colnumbers
\startdata
\multirow{2}{*}{\shortstack[l]{Single \\Power law}} & Slope & $-1.45$ & $[-1.75, -1.16]$ & $[-2.11, -0.88]$ & $[-2.56, -0.68]$ \\
							      & Binary Fraction & 0.25 & $[0.0, 0.33] $& $[0.0, 0.58]$ & $[0.0, 0.83]$ \\
\hline
\multirow{4}{*}{\shortstack[l]{Broken \\Power law}} & Low-mass Slope   & $-1.18$ & $[-1.51, -0.69]$ & $[-1.93, -0.49]$ & $[-1.97, -0.11]$\\
                                                    & High-mass Slope  & $-1.88$ & $[-2.37, -1.45]$ & $[-2.66, -1.00]$ & $[-2.95, -1.00]$\\
                                                    & Break mass       &  0.56 & $[0.44, 0.68] $  & $[0.36, 0.79] $  & $[0.30, 0.80]$\\
							                        & Binary Fraction  &  0.23 & $[0.07, 0.35]$    & $[0.00, 0.50]$    & $[0.00, 0.86]$\\
\hline
\multirow{3}{*}{Log-normal}       & $m_c$ & 0.33 & $[0.16, 0.49] $& $[0.03, 0.65]$ & $[0.02, 0.86]$\\
                                  & $\sigma$& 0.68 & $[0.56, 0.84]$ & $[0.35, 0.95] $& $[0.23, 0.99]$\\
							      & Binary Fraction & 0.20 & $[0.0, 0.26]$ & $[0.0, 0.53]$ & $[0.0, 0.96]$\\
\enddata
\end{deluxetable*}
\section{Results}
\label{sec:results}

Given that both the photometry and the artificial star tests have been performed on each of the 4 tiles separately, we perform MCMC experiments on the individual tiles, and then combine the posteriors to obtain the average properties of the Com~Ber IMF across the tiles. As in Paper~I, the MCMC experiments are performed assuming a flat prior on the intensity, i.e. the total number of expected stars. 
This easily allows choosing a different prior, by using a reweighting of the samples.
\begin{comment}
Specifically, for our final results we adopt a logarithmic prior on the intensity, which is achieved by reweighting individual samples by intensity$^{-1}$. 
The logarithmic prior choice is well justified: it is an observed fact, as well as a prediction of cosmological models, that the Universe contains more small galaxies than large ones. Therefore, a priori, smaller galaxies, containing a lower number of stars, are preferred.
\end{comment}
Specifically, for our final results we adopt a uniform prior on the logarithm of the intensity, which is achieved by reweighting individual samples by intensity$^{-1}$.
This prior choice is well justified: it is an observed fact, as well as a prediction of cosmological models, that the Universe contains more small galaxies than large ones. 
For example, in the \cite{1976ApJ...203..297S} galaxy luminosity function, at least well below the cutoff luminosity $L^*$, the dependency of the number of galaxies with their luminosity (of which the number of stars within a galaxy is a proxy) is that of a power-law with index close to $-1$.
Therefore, a priori, smaller galaxies, containing a lower number of stars, are preferred.

\begin{figure*}
\gridline{\fig{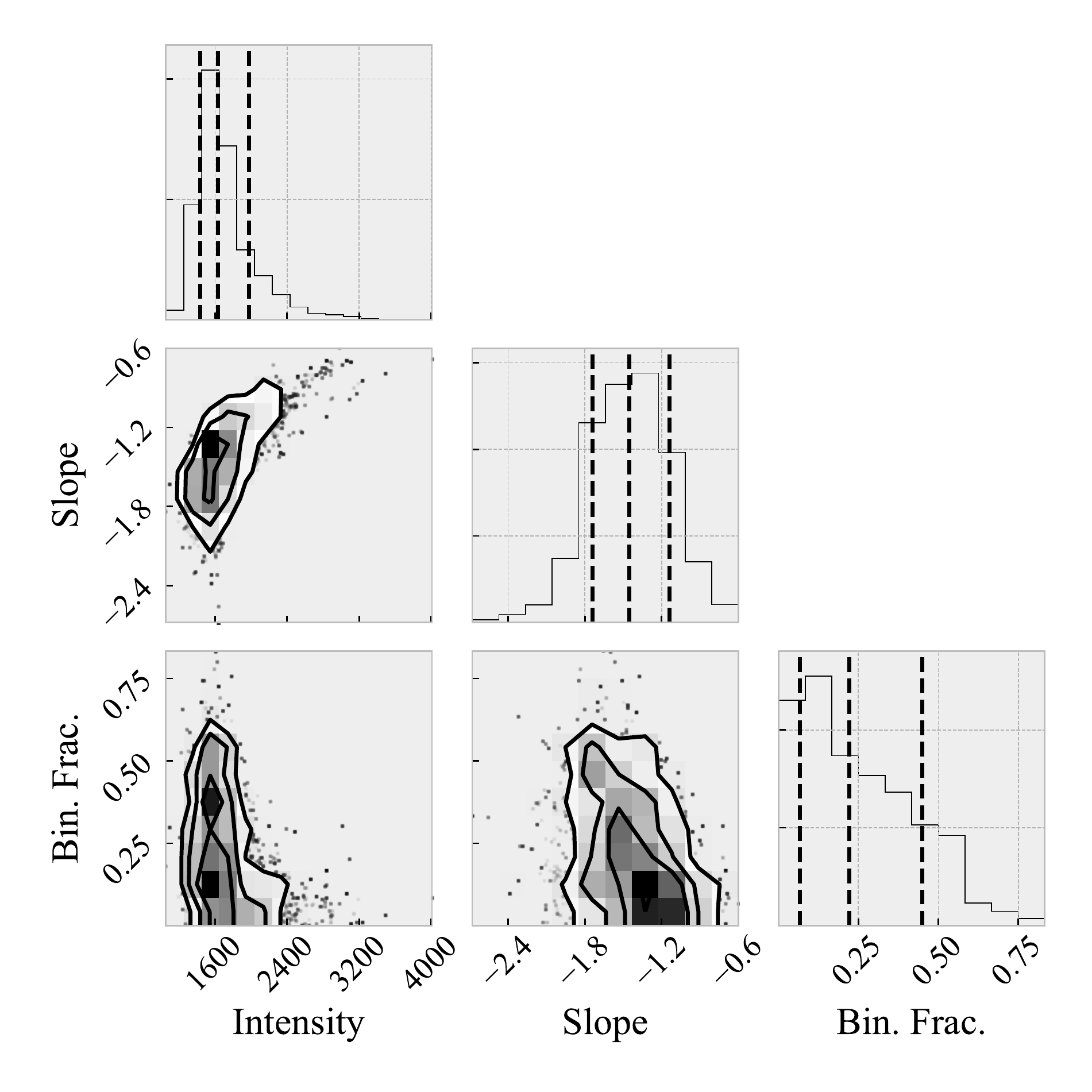}{.39\textwidth}{Single Power Law}
		  \fig{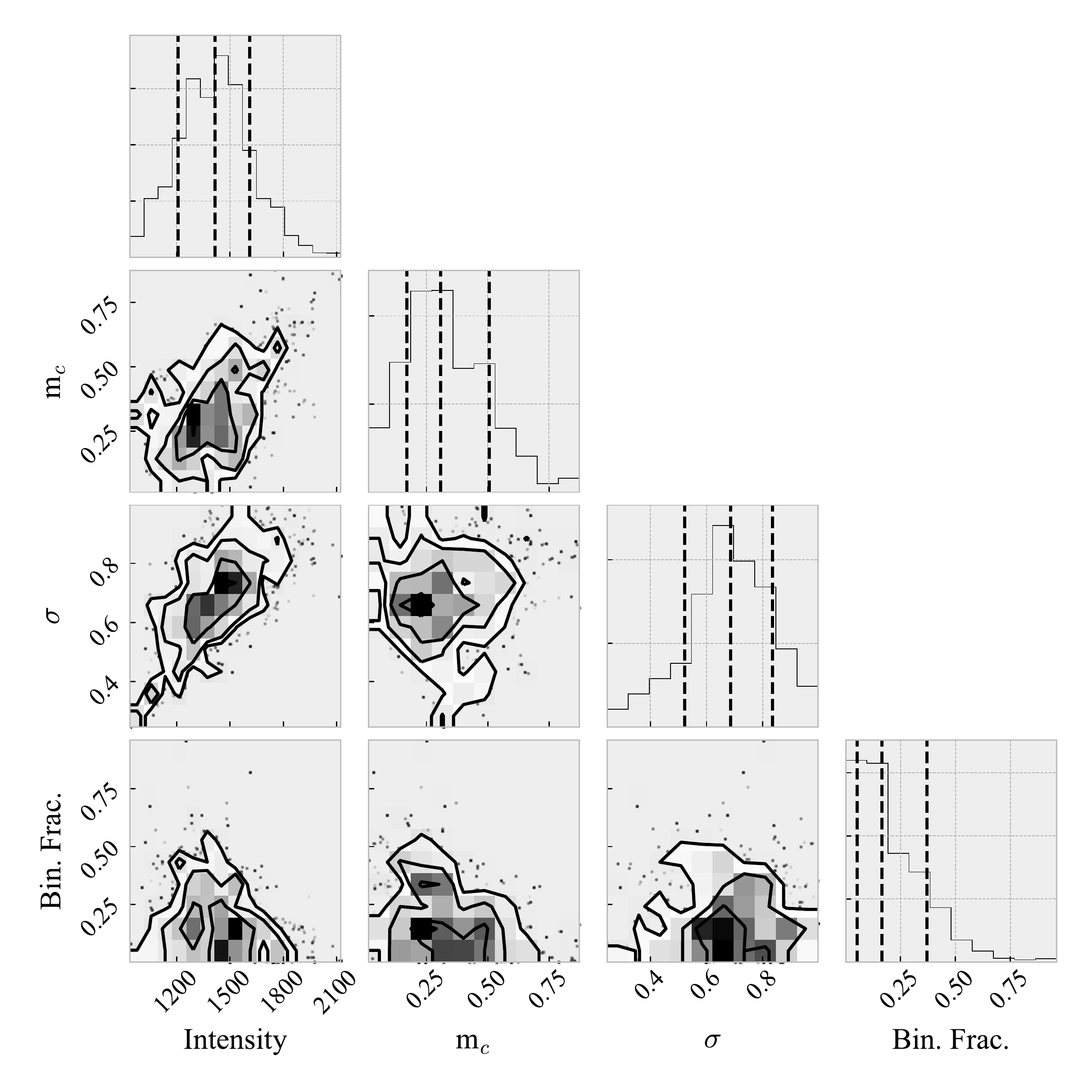}{.52\textwidth}{Log-normal}}
\gridline{\fig{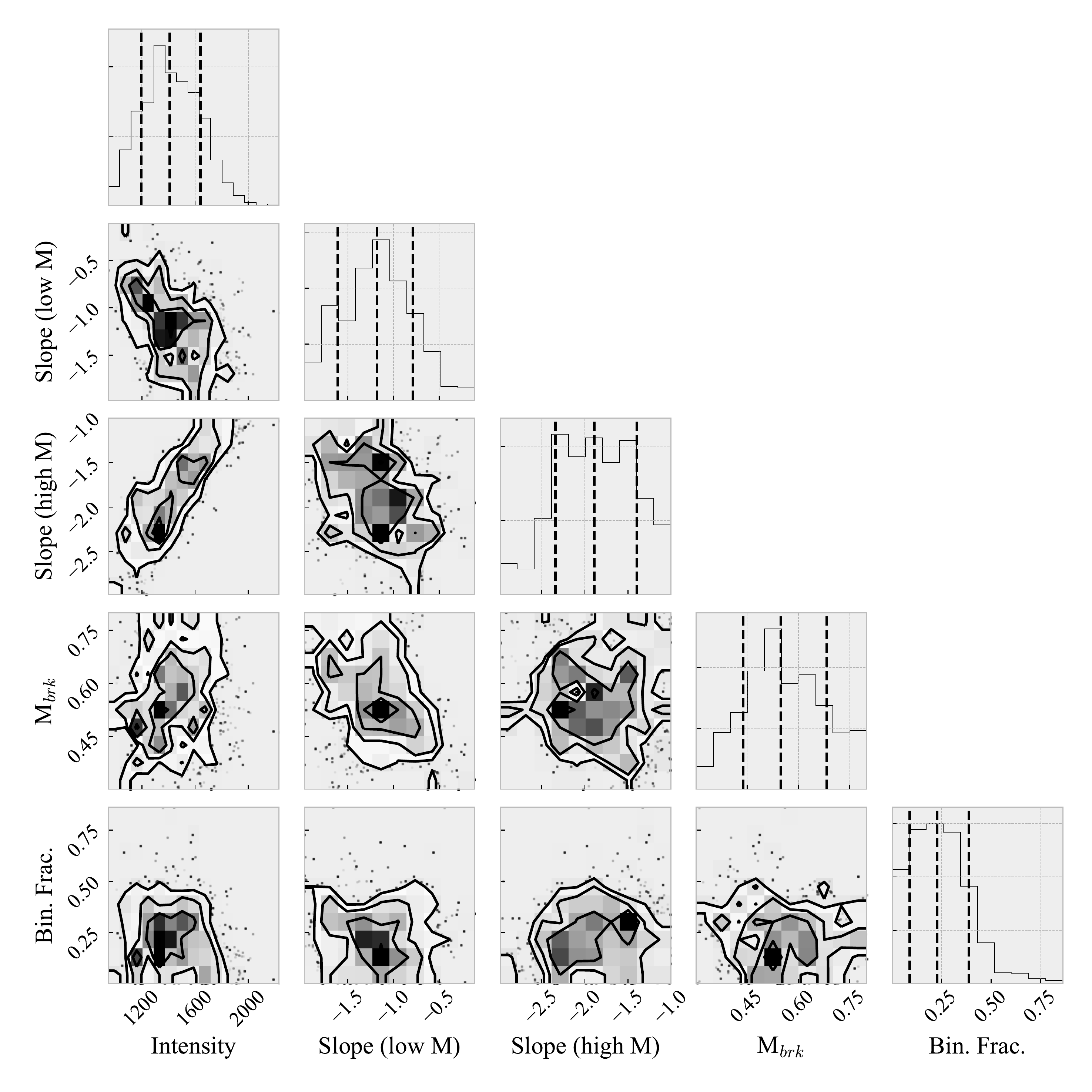}{.65\textwidth}{Broken Power Law}}
\caption{Corner plots of the posterior distributions of IMF parameters for the 3 different models. The vertical dashed lines indicate the 16, 50 (median), 84 percentiles of the marginal distributions. \label{fig:cp}}
\end{figure*}

The results for the 4-tile-combined posteriors of the IMF parameters, for the three different model choices (Single Power Law, Broken Power Law, Log-normal; hereafter SPL, BPL and LN, respectively) are displayed in Figure~\ref{fig:cp}, while the best-fit values and credible intervals are summarized in Table~\ref{tab:res}.
We use the mean of the marginal posterior of each parameter as the best-fit estimator. Our credible intervals are defined as the smallest intervals containing the 68\%, 95\% and 99\% of the marginal probability, we sometimes refer to these as 1$\sigma$, 2$\sigma$ and 3$\sigma$ intervals respectively.

\begin{figure*}
\gridline{\fig{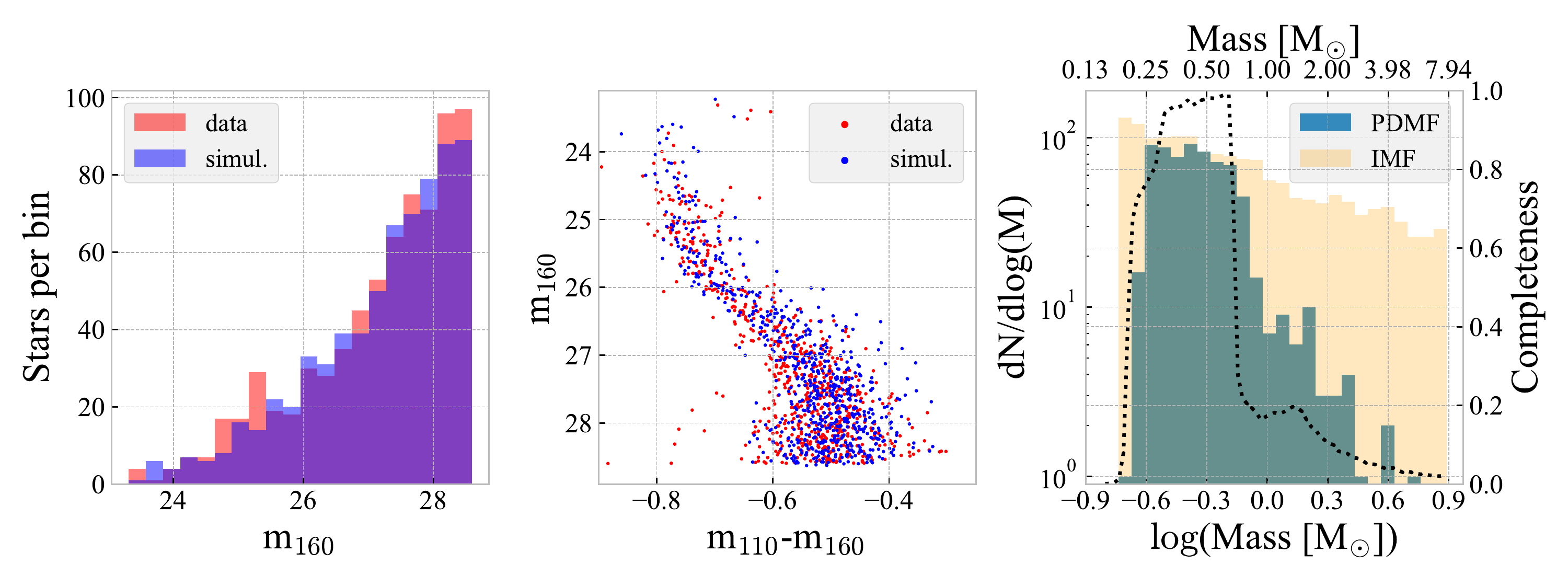}{.88\textwidth}{Single Power Law}}
\gridline{\fig{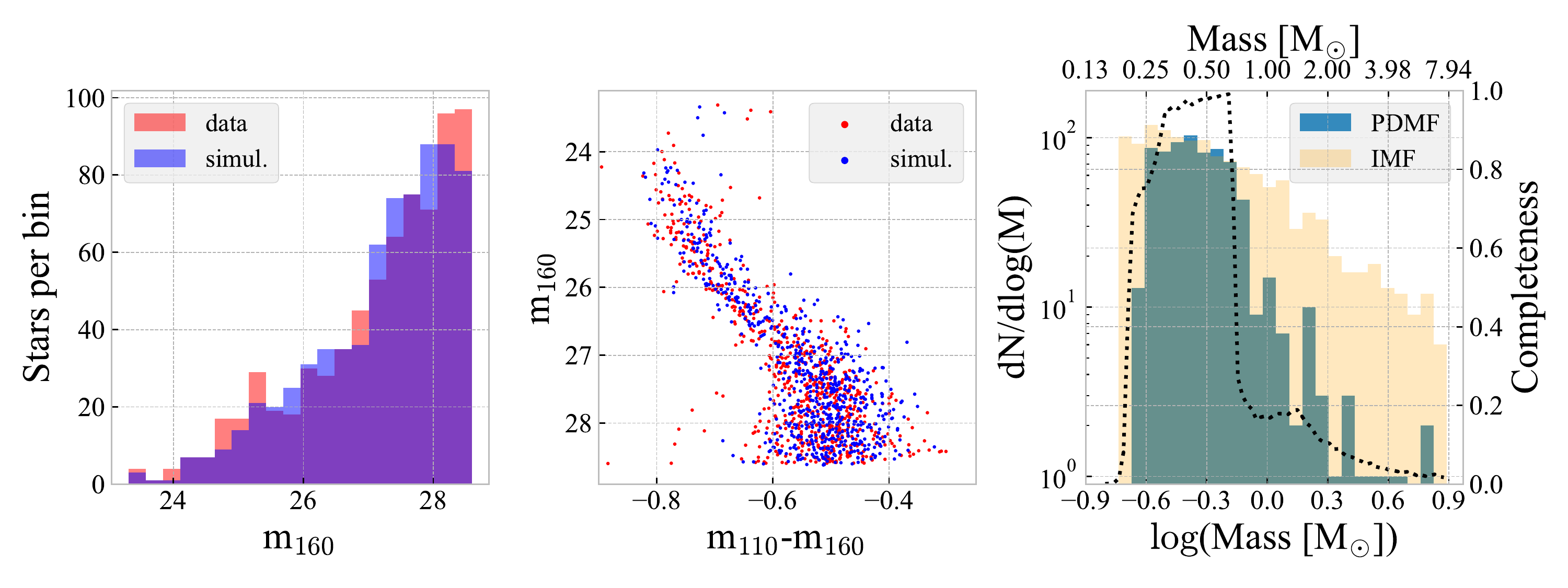}{.88\textwidth}{Broken Power Law}}
\gridline{\fig{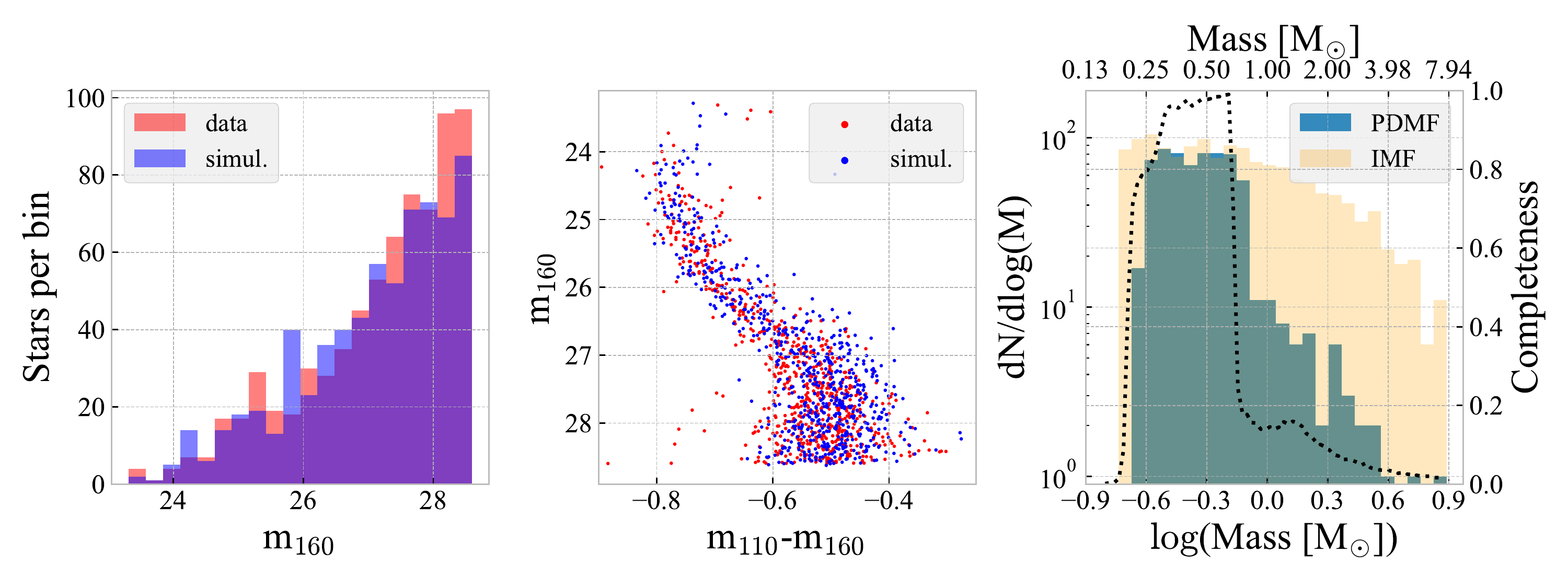}{.88\textwidth}{Log-normal}}
\caption{Comparison between the simulations (blue) for the best-fit parameters of each model and the data (red). Left: simulated and observed luminosity function in the F160W band. Center: simulated and observed CMD. Right: draws from the underlying IMF (yellow/orange), and corresponding Present Day Mass Function (PDMF, teal), i.e., the mass distribution of the stellar systems that are actually observed in the simulated CMD. The dashed lines show the ratio of the two, i.e., the completeness as function of system mass; to obtain the dashed line a much higher number of stars have been drawn than those shown here. Note that the PDMF extends beyond the turnoff mass because there are binary systems of total mass larger than the turnoff mass, where only the secondary star, less massive than such limit, survives. \label{fig:bestfits}}
\end{figure*}

Figure~\ref{fig:bestfits} shows a comparison between synthetic CMDs and luminosity functions computed for the best fit parameter sets and the data.
The right panels of this Figure show the difference between the complete set of extracted masses (the IMF, orange) and the masses of the objects that are ``observed'' in the synthetic CMD (the present day mass function, PDMF). 

\section{Discussion}

%Regardless of the adopted functional form of the IMF, the findings are unequivocal: the IMF parameters for Coma Berenices differ significantly from those of the Milky Way. Summarizing the results from the 3 models, we can say that i) Coma Berenices contains relatively less low-mass stars (below 0.5~M$_{\odot}$) with respect to the Galaxy, ii) the mean stellar mass for stars in the probed mass range, $0.2 \lesssim \mathrm{M}_{\odot} \lesssim 0.8 $, is larger than that in the Galaxy.
%Given the limited mass range and the small number of stars (717), the significance of our results is limited. For example, the standard Salpeter value of -2.3 for the slope of a single power law falls outside our 95\% credible interval, but inside our 99\% one.
%Similarly we can rule out the Chabrier (Galactic) values for a log-normal IMF only at the 95\% credibility level.
%The broken power-law parameters we obtain are the only ones that differ form the accepted Galactic model (a Kroupa IMF) at more than a 99\% level.

\subsection{Comparison between the different models}
From Figure~\ref{fig:bestfits} we can see that the three IMF models give very similar results. The $m_{160}$ luminosity functions for the best fit parameters for each of the three models (left columns in the figure) do trace the observed luminosity function, even if with some differences in the details. %Given the limited size of our sample of stars, and the systematic uncertainties from the stellar models, it is however beyond the scope of this paper to try and interpret these details as favoring one form of the IMF over the other. We thus do not attempt a quantitative assessment of the quality of fit nor perform any quantitative model comparison. 
From the right panels of the Figure~\ref{fig:bestfits}, the best-fit parameters for the three IMF models produce very similar PDMFs (green), regardless of the different underlying IMFs (orange). We note some discrepancy between the models, especially at the high-mass end, where there are the fewest stars.
Another fact emerges quite clearly: no obvious turnover of the IMF is observed. The SPL model gives as good a fit to the data down to our limiting mass as the BPL and LN ones; if anything the best-fit SPL model seems to be the one producing the luminosity function giving the best qualitative agreement at faint magnitudes.

To provide a more quantitative assessment of the goodness of fit (gof) and see whether any of the 3 models is inconsistent, in a gof sense, with the data, we devised an ad-hoc strategy.
Our fitting method uses an ABC approach, and thus bypasses an explicit calculation of the likelihood. Methods like Bayes Factors, Akaike- or Bayesian Information criteria (AIC, BIC) cannot therefore be used here.
Our acceptance of a proposed parameter set, $\vec{\theta_i}$ relies on generating a CMD using $\vec{\theta_i}$, which we call $y_i$ and determine whether $y_i$ is close to the observed CMD, $y_{obs}$.
The distance between the simulated and observed CMD is given by equation (10) of \cite{2018ApJ...855...20G}; we will call such distance $\rho(y_i,y_{obs})$.

Our gof criterion is a modification of the gof 
statistic based on the posterior predictive distribution proposed by \cite{2016arXiv160104096L}, and it consists of:
\begin{itemize}
\item take two subsets of draws from the posterior, 
a training set $\{\vec{\theta_j}\}_{j\in J}$, and a test set $\{\vec{\theta_k}\}_{k\in K}$ possibly from two disjoint parts of the MCMC chain. In practice we use the first 250 draws for $J$ and the last 1000 for $K$
\item compute, for each $k\in K$, the average distance of the corresponding CMD, $\vec{\theta_k}$ to the $\{\vec{\theta_j}\}_{j\in J}$:
\begin{equation}
D_{post,k} = \frac{1}{n_J}\sum_{j\in J}\rho(y_k,y_j),
\end{equation}
where $n_J$ is the size of the training set.
\item similarly compute the average distance to the training set for the observed CMD:
\begin{equation}
D_{post,obs} = \frac{1}{n_J}\sum_{j\in J}\rho(y_{obs},y_j)
\end{equation}
\item see where $D_{post,obs}$ falls within the distribution of $D_{post}$ computed from the test set
\item if the average distance for the observations falls outside some pre-defined quantile (e.g. if the observations are on average more distant than 95\% of the $\{\vec{\theta_k}\}_{k\in K}$), rule out that model
\end{itemize}

\begin{figure*}
\includegraphics[width=0.33\textwidth]{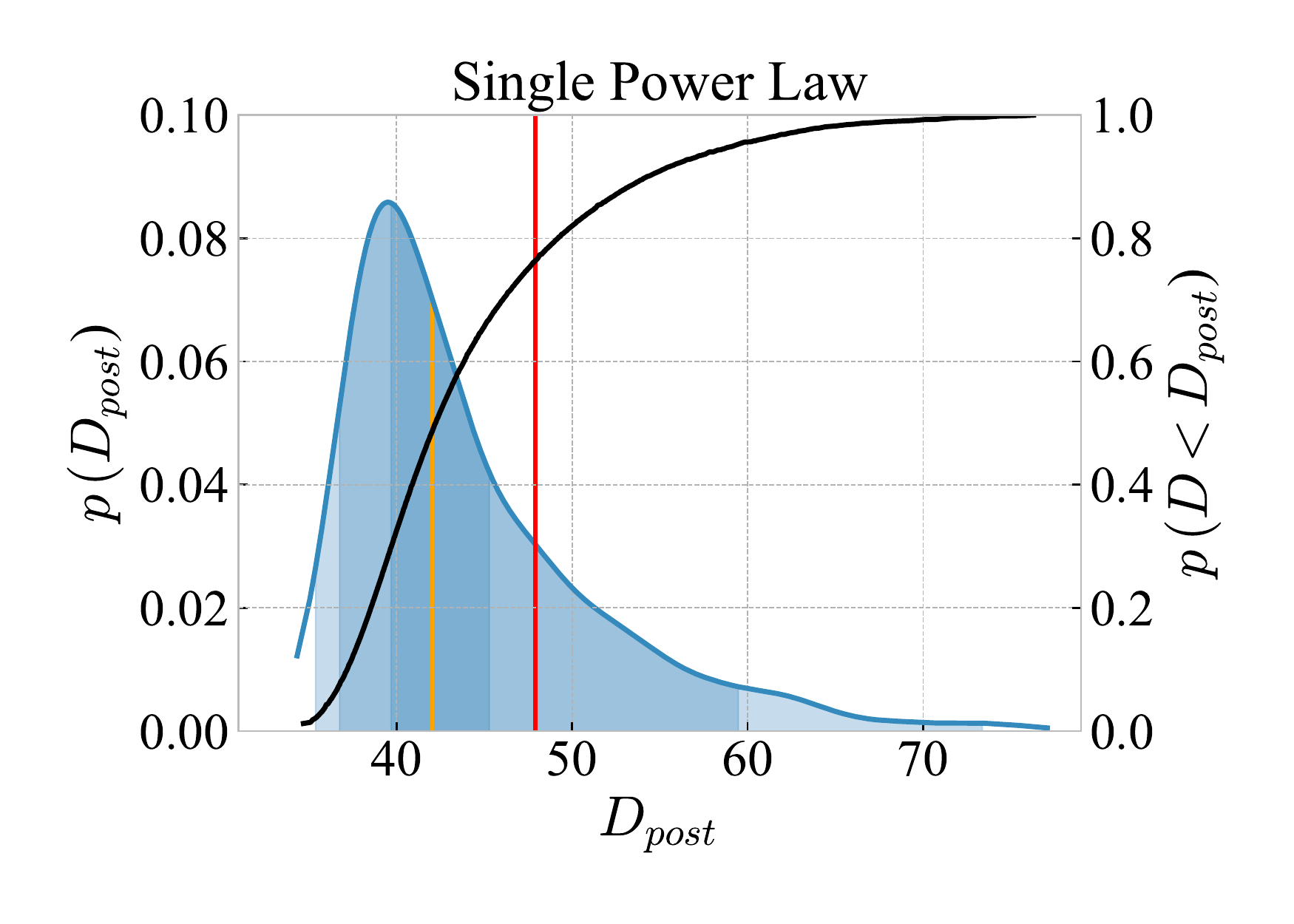}
\includegraphics[width=0.33\textwidth]{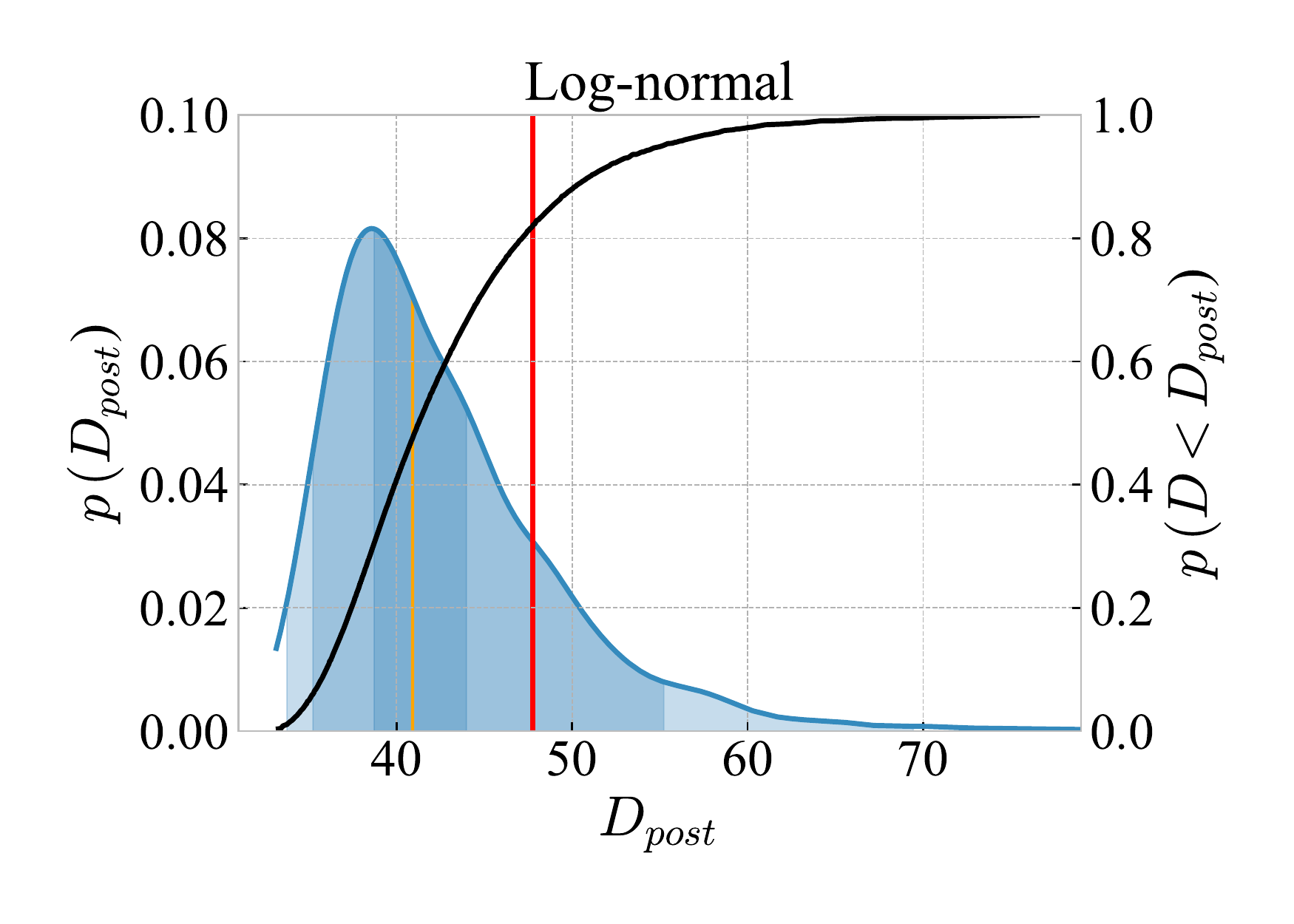}
\includegraphics[width=0.33\textwidth]{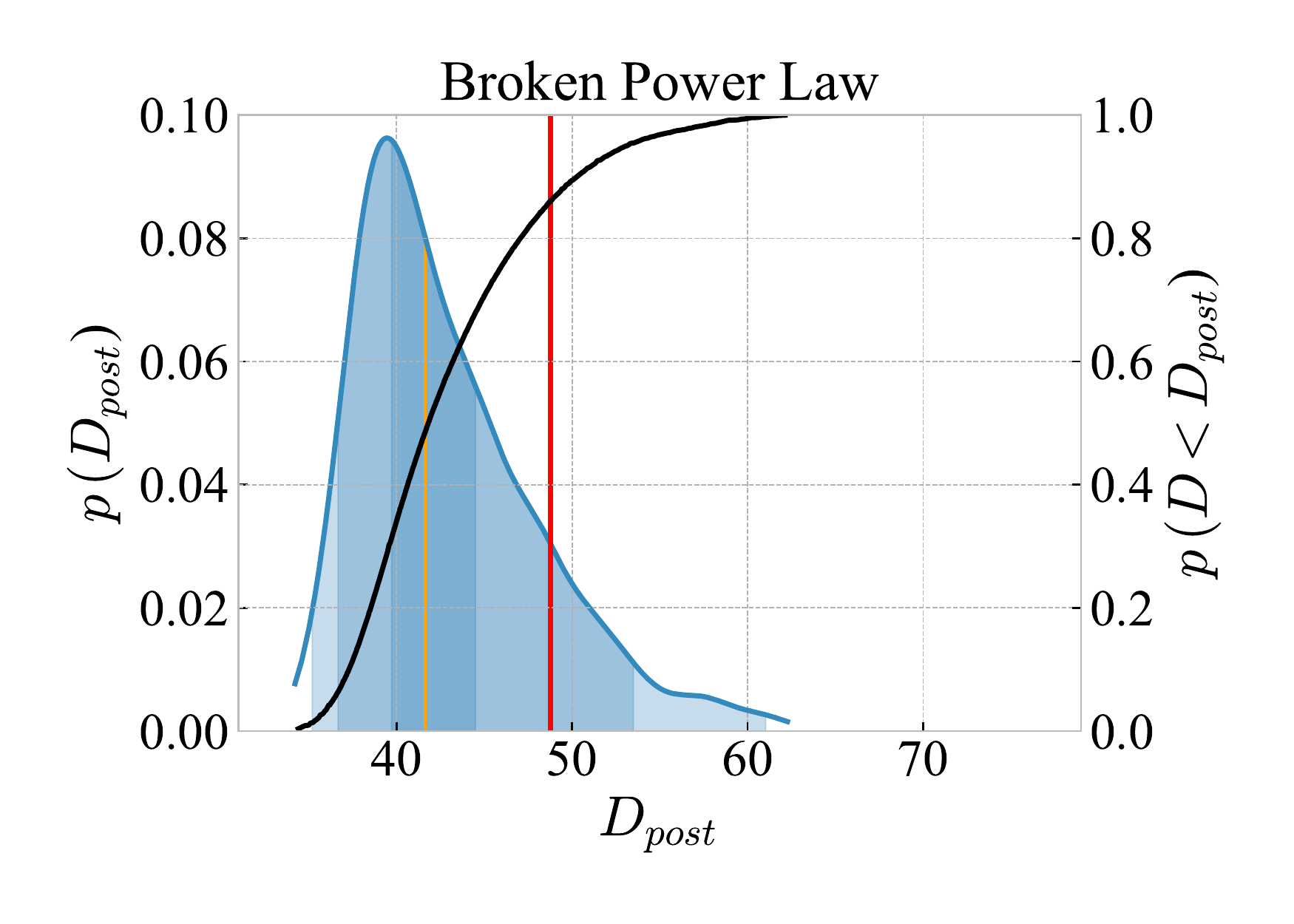}
\caption{Goodness of fit test based on the posterior predictive distribution, $D_{post}$, for each of the three adopted IMF models.
The shades of blue in the distributions indicate the 68\%, 95\% and 99\% quantiles, the orange vertical line the median.
The black line is the cumulative (right y-axis of each panel). The red line indicates the value of $D_{post}$ for the observed dataset. Since these values are always withing the 95\% interval, no model can be rules out by our goodness of fit test.
\label{fig:gofs}}
\end{figure*}

Figure~\ref{fig:gofs} shows the results for each of the three IMF models. In each case the $D_{post,obs}$ falls well within the bulk of the distribution of $D_{post}$. For none of the 3 models the fit is outstandingly better or worse than for the others.
All three models are not inconsistent with the data according to our gof test. This confirms the qualitative conclusion that could be drawn from the luminosity functions of Figure~\ref{fig:bestfits}: all three IMF models can produce CMDs that are consistent with the observed CMD.

%Both the log-normal and broken power law models suggest that the characteristic IMF mass is slightly higher than the Milky Way value. It is indeed possible that a turnover exists, but at masses above $\sim0.6\, \mathrm{M}_{\odot}$. It is however hard to test this interesting hypothesis with the data in hand, for two main reasons. First, the already low total number of stars would be further reduced if we considered only stars more massive than 0.6~$\mathrm{M}_{\odot}$. Second, the limited mass range available at masses above the supposed turn-over (the main sequence turn-off mass is about 0.8~$\mathrm{M}_{\odot}$) may hamper a clear determination of such a feature even if more stars were available.

\subsection{Comparison with results from optical data}
\label{sec:comp_opt}
\begin{figure*}
\includegraphics[width=0.495\textwidth]{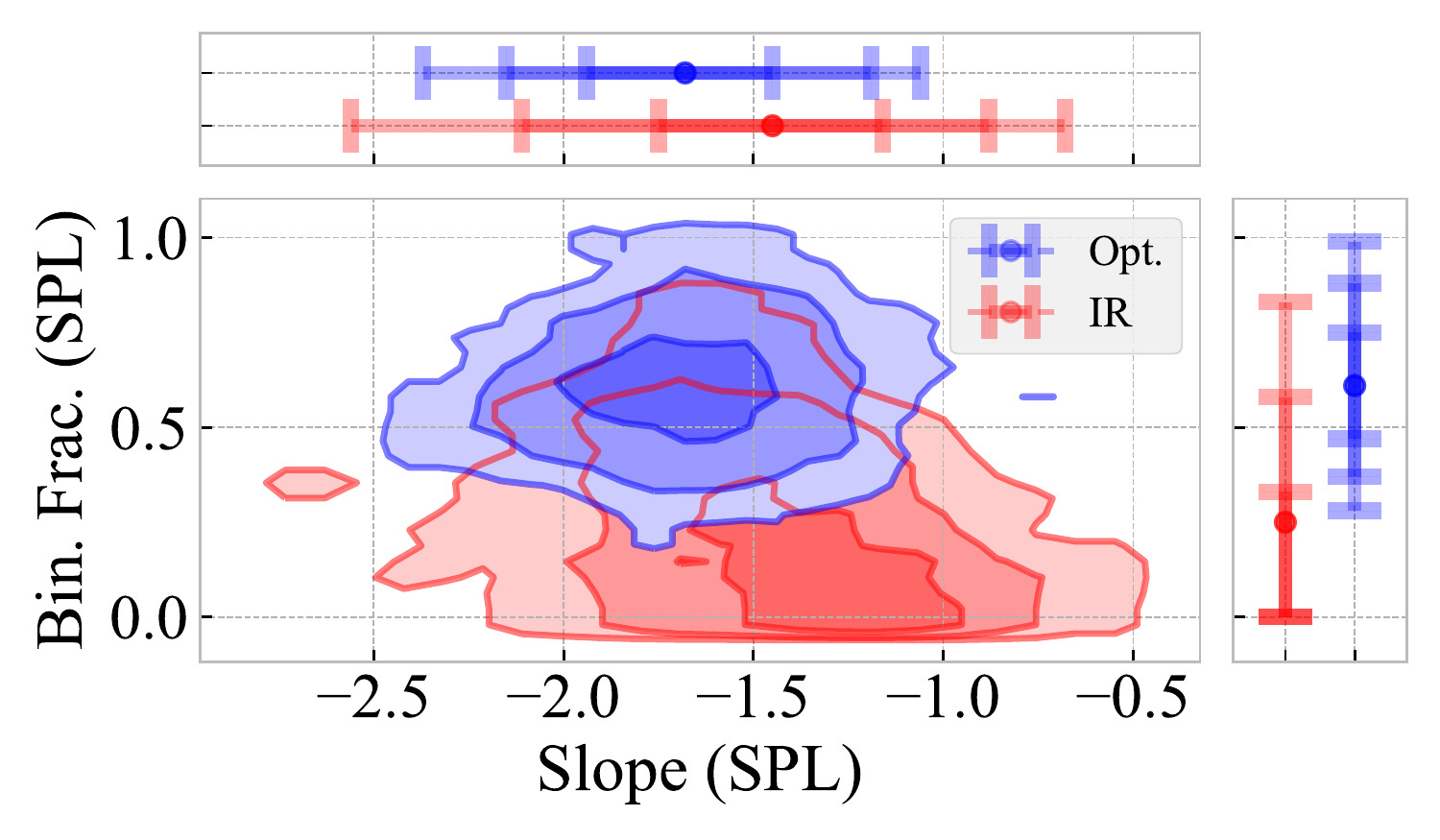}
\includegraphics[width=0.495\textwidth]{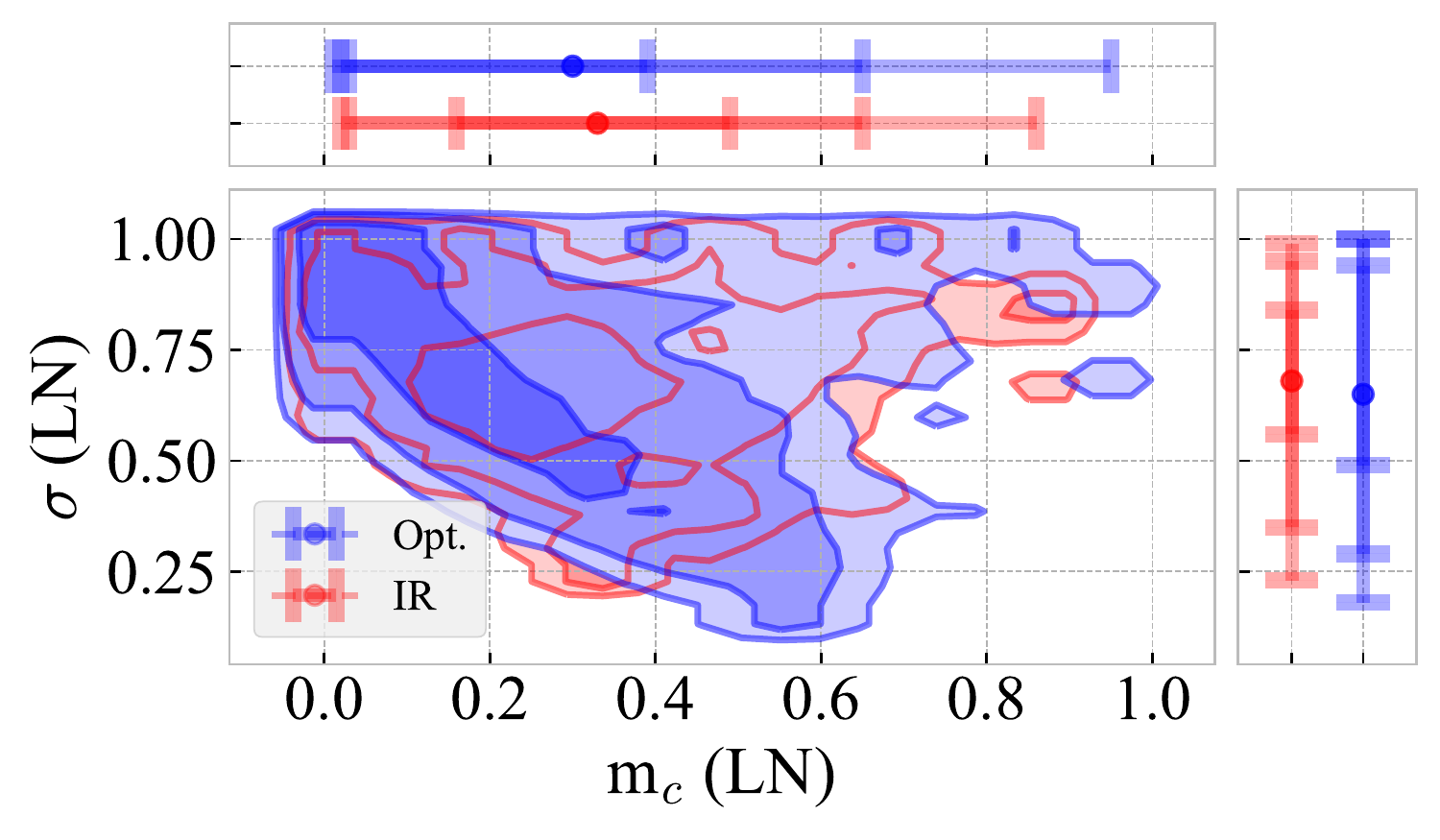}
\includegraphics[width=0.495\textwidth]{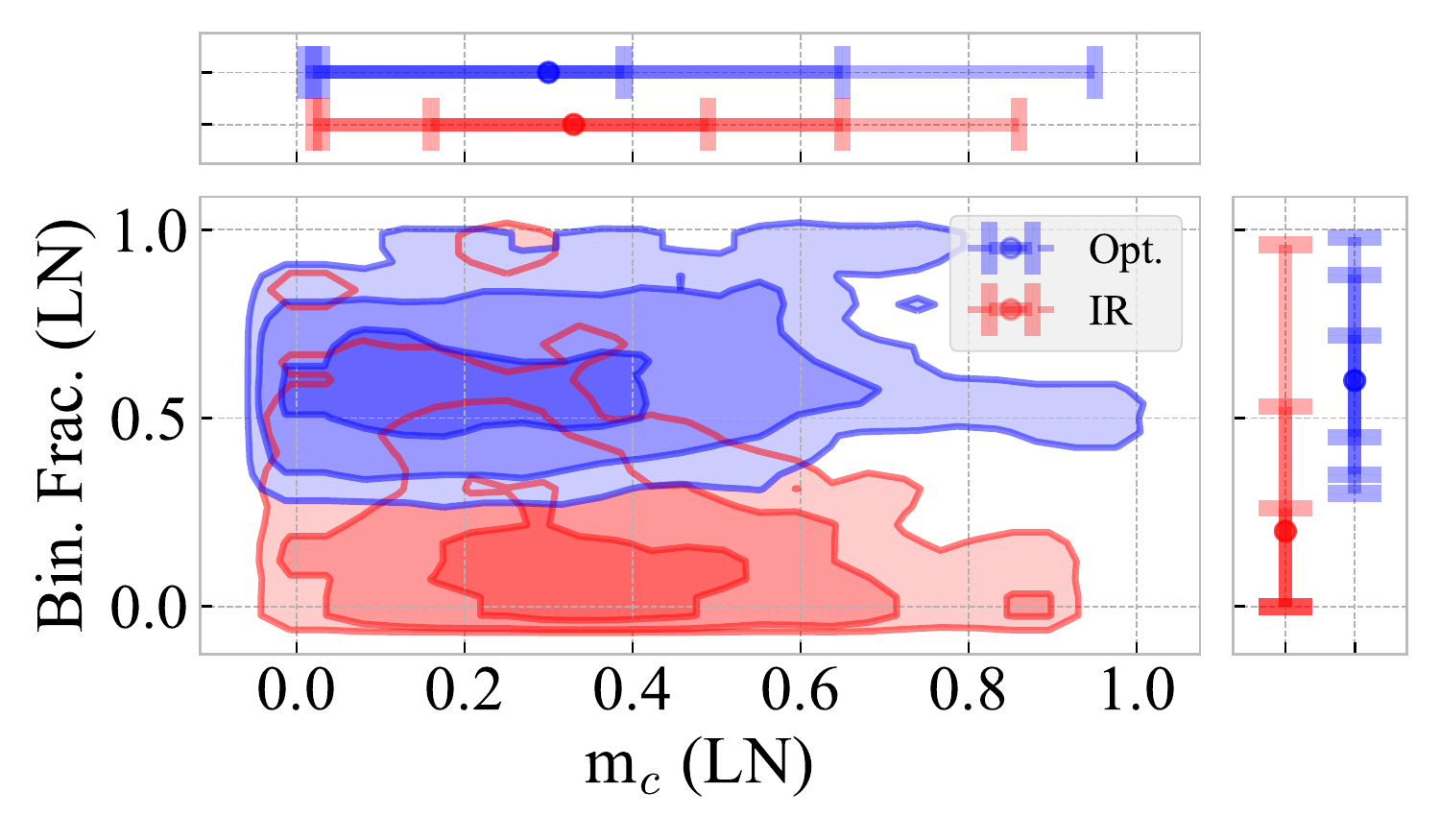}
\includegraphics[width=0.495\textwidth]{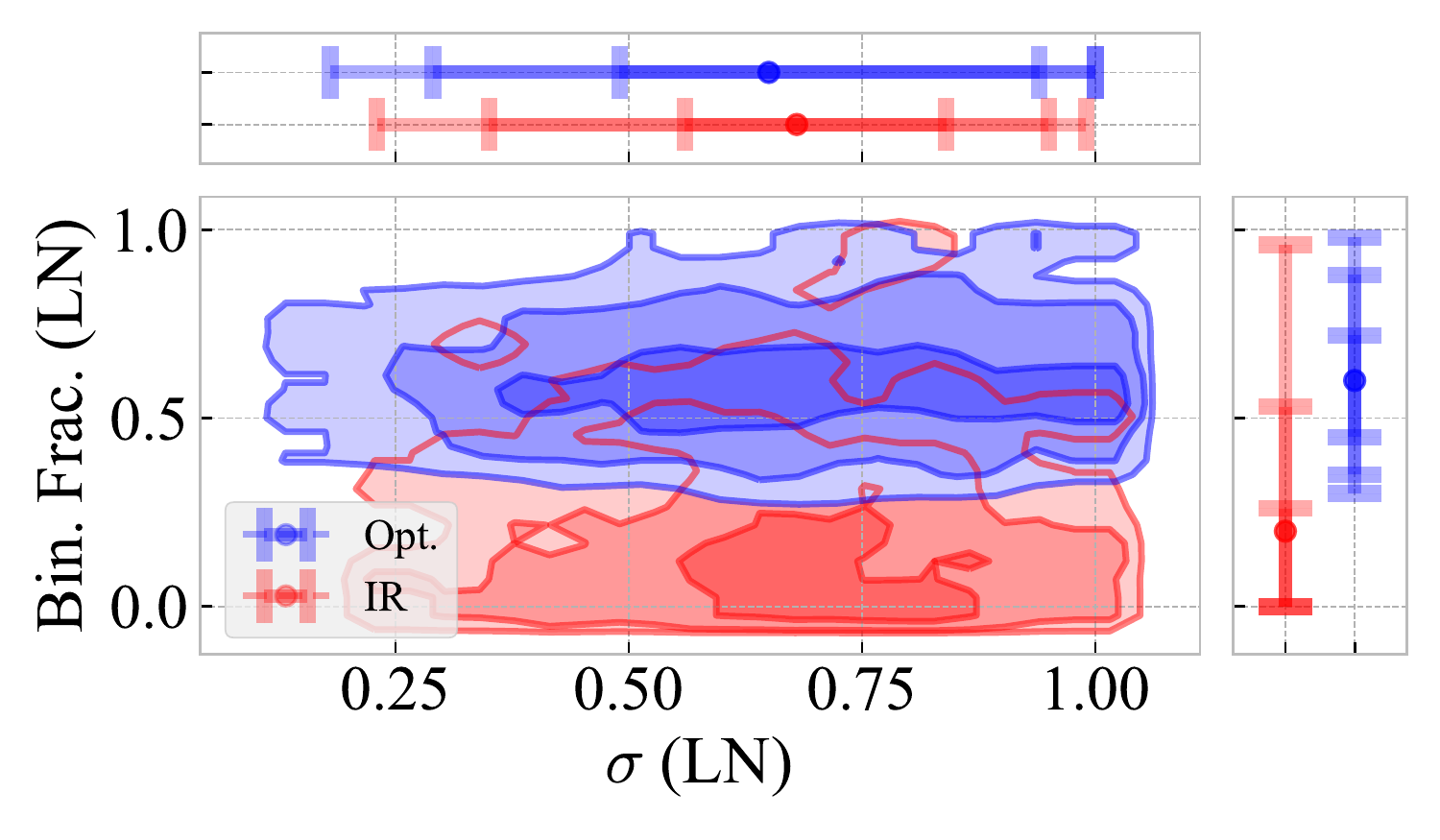}
\caption{Comparison of the results for the optical data of ComBer \citep[][, blue]{2018ApJ...855...20G} and the IR data of the present paper (red). The contour plots show pairwise relations for the Single Power Law (SPL) model parameters, and the Log-Normal (LN) model parameters. The top and left plots in each quadrant show the 68-95-99 \% credibility intervals ranges for each parameter, obtained from the respective marginal probability distributions\label{fig:compareOPTIR}.}
\end{figure*}

When compared to the results of \cite{2018ApJ...855...20G}, our current results for the SPL and LN model IMF are in good agreement, namely the SPL slope and the LN $m_c$ and $\sigma$ are within the 68\% C.I. (see Figure~\ref{fig:compareOPTIR}). 
This is an important sanity check for both works, demonstrating that any bias in the results due to the different mass ranges probed, as well as to uncertainties in the adopted color-temperature and/or the mass-luminosity relations, if present, must be negligible with respect to the uncertainties in the parameters estimate.

We cannot perform a comparison for the broken 
power-law model, since \cite{2018ApJ...855...20G} did not attempt a BPL modeling of the IMF to fit their optical data. This was because their data only reach down to $\sim0.4 - 0.5 \mathrm{~M}_{\odot}$ on average for the six galaxies of their sample.

The binary fractions are generally in worse agreement between the current work and \cite{2018ApJ...855...20G}.
The derived binary fraction is quite sensitive to the observed width in the CMD. As illustrated in section \ref{sec:extraerror}, we were forced to add an extra error in order for our simulated CMDs to reproduce the observed CMD width. We used the same extra error in each band and at all magnitudes, because a more detailed model for it would have implied a better ``a priori'' knowledge of what the width of the simulated CMD should have looked like, i.e., a knowledge of the binary fraction, among other things.
We speculate that the necessarily qualitative-only assessment of the extra errors may have caused an extra spread of the simulated main sequence, which in turn translates into a smaller binary fraction, to compensate for the extra width.
On the other end, in \cite{2018ApJ...855...20G}, Com~Ber was the only UFD with a best-fit binary fraction higher than 50\%, while for the other UFDs the best fit binary fraction was always smaller than 30\%. Thus it could be that the binary fraction was overestimated in that work as well.
We do not try to draw further conclusions from a discrepancy between the derived binary fractions.

\subsection{Comparison with the Milky Way}

If limited to the SPL parametrization, our results would suggest that the mean IMF slope of Com~Ber is somewhat shallower from the Milky Way classical Salpeter slope of $-2.35$.
At the same time, given the limited mass range and the small number of stars (717), the significance of such a result is limited. The Salpeter value falls outside the 95\% credible interval, but inside the 99\% one.

The BPL and LN parametrizations give results that are more similar to the Milky Way  for the same type of functional forms, i.e. \cite{2001MNRAS.322..231K} for the BPL and \cite{2003PASP..115..763C} or \cite{2010AJ....139.2679B} for the LN form.
While the best-fit parameters for both the BPL and LN model would result into a slightly bottom lighter IMF for Com~Ber with respect to the Galaxy, the \cite{2001MNRAS.322..231K} and \cite{2003PASP..115..763C} values are within the 68\% confidence interval.

\cite{2018ApJ...855...20G}, based on slope values from SPL model fits, claimed that, on average, a group of 6 UFDs showed a more bottom light IMF with respect to the Milky Way, i.e. fewer dwarf stars. 
It is possible that the differences between the IMF for the UFDs with respect to the Galaxy claimed by \cite{2018ApJ...855...20G} are, at least partly, an artifact of using a SPL model.
\cite{2018ApJ...855...20G} studied the mass range above 0.4-0.5 M$_{\odot}$, at the edge where \cite{2001MNRAS.322..231K} place the power-law break, and claimed to not expect a significantly biased result for an SPL model.
On the other hand, \cite{2017MNRAS.468..319E} show how adopting a SPL model could artificially lead to shallow slopes if the true underlying IMF has a log-normal shape and data do not probe down to the characteristic mass. \cite{2017MNRAS.468..319E} obtain a slope of $\sim -1.55$ by fitting a \cite{2003PASP..115..763C} IMF with an SPL model, but limiting the fit to the $(0.4 - 0.77)$~M$_{\odot}$ range, similar to what obtained by \cite{2018ApJ...855...20G} for Com~Ber.
The fact that the results of the current work, for the SPL model, using an extended mass range which reaches very close to the \cite{2003PASP..115..763C} characteristic mass, are very similar to \cite{2018ApJ...855...20G} shows however that any bias, if present, must be smaller than the uncertainty introduced by the limited number of stars used.

It is worth noting that in \cite{2018ApJ...855...20G} Com~Ber was, together with Boo~I, the UFD for which the IMF best-fit parameters for both an SPL and LN model were closer to those of the Milky Way. Other galaxies in \cite{2018ApJ...855...20G}, in particular Hercules, Leo~IV and CVn~II, showed a much more significant discrepancy, and mean slope values close to $-1$. Also the characteristic mass for the LN model for these three galaxies was much higher than the \cite{2003PASP..115..763C} value, thus for those galaxies the discrepancy with the Milky Way was present even when using the LN parametrization (although, again with lower significance than when using an SPL model).

\cite{2018ApJ...855...20G} noted that Com~Ber and Boo~I, being the closest UFDs in their sample, subtended a larger angle in the sky, and thus their CMDs could have been more contaminated by unresolved background galaxies, specially at the faint end, making their IMF artificially steeper.
In the current work we have tried to minimize the impact of background interlopers by limiting ourselves to $m_{160} < 28.6$ mag, where we can still discriminate between extended background galaxies and point sources.
This does not ensure 100\% purity in our sample, and thus may be part of the reason for having an IMF that does not differ from the Milky Way as significantly as for other UFDs.

%Finally it is possible that the IMF of Com~Ber is truly consistent with the Milky Way one, and that using a SPL parametrization is simply an incorrect way to perform a comparison.
%Given the qualitative agreement of all 3 paramterizations used in the current work, an given that no strong bias is evident in the inferred SPL slope, 
%and in light of the observed variance in slope between the 6 UFDs of Gennaro et al. (2018), it is possible that the SPL parametrization can still be safely applied in the mass range of our IR data for Com~Ber, as well as the optical data of Gennaro et al. (2018).

\subsection{The effect of fixing the binary fraction}

In Section~\ref{sec:comp_opt} we illustrated some possible concerns on the reliability of the estimated binary fraction, related to the extra error added to our artificial star tests in order to reproduce the observed width of the CMD.
With those caveats in mind, it is still interesting to dissect the multi-dimensional posterior probability distribution functions in order to emphasize the correlation of the other parameters with the binary fraction itself.

For each of the SPL, LN and BPL models, we have divided the posterior draws in 4 parts, corresponding to the 0 to 25\%, 25\% to 50\%, 50\% to 75\% and 75\% to 100\% quantiles of the marginal distribution of the binary fraction. The edges of the quantiles change slightly between the different models, according to the detailed difference in the marginal distribution of the binary fraction.

\begin{figure*}
\includegraphics[width=0.34\textwidth]{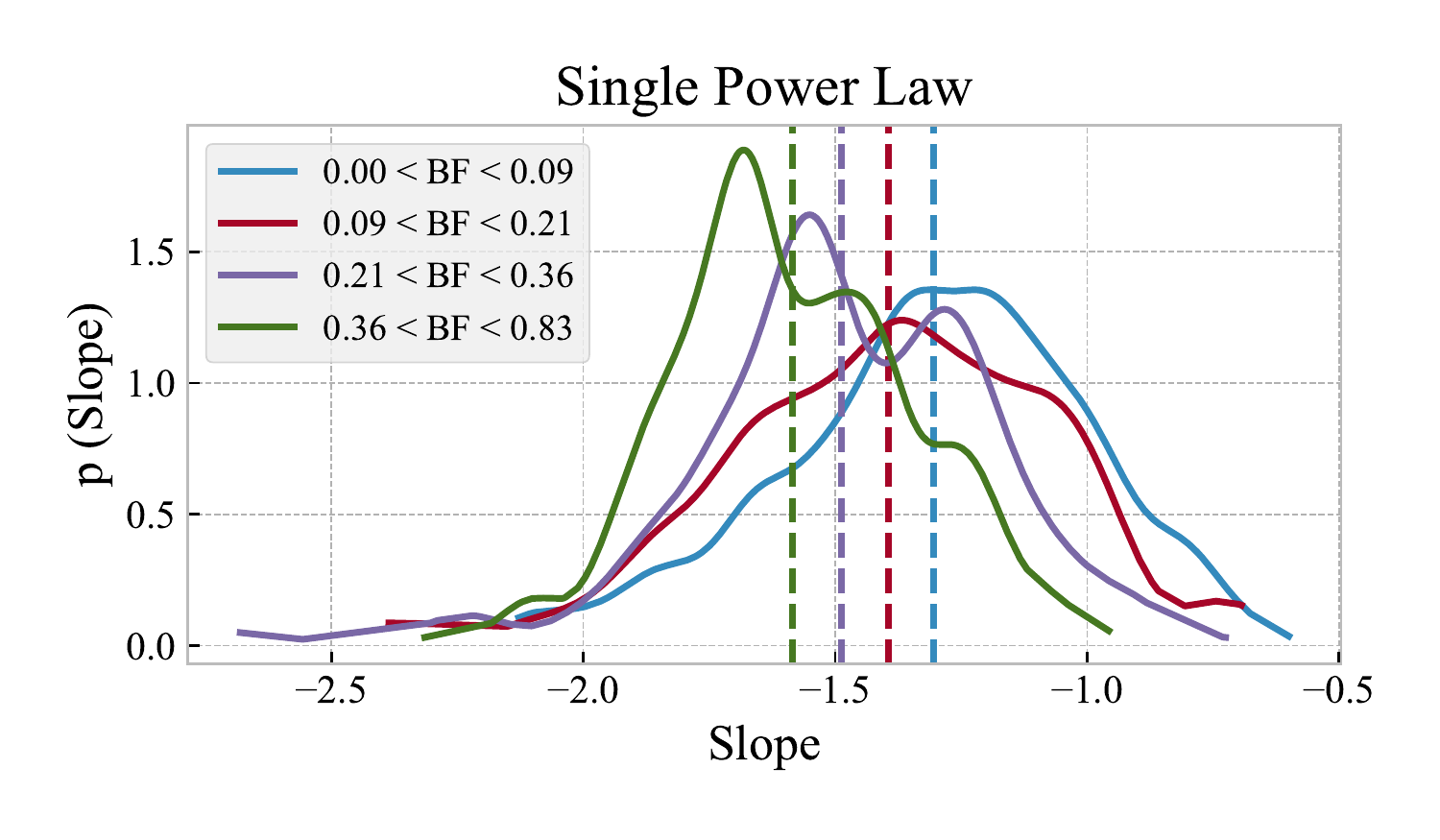}\\
\includegraphics[width=0.67\textwidth]{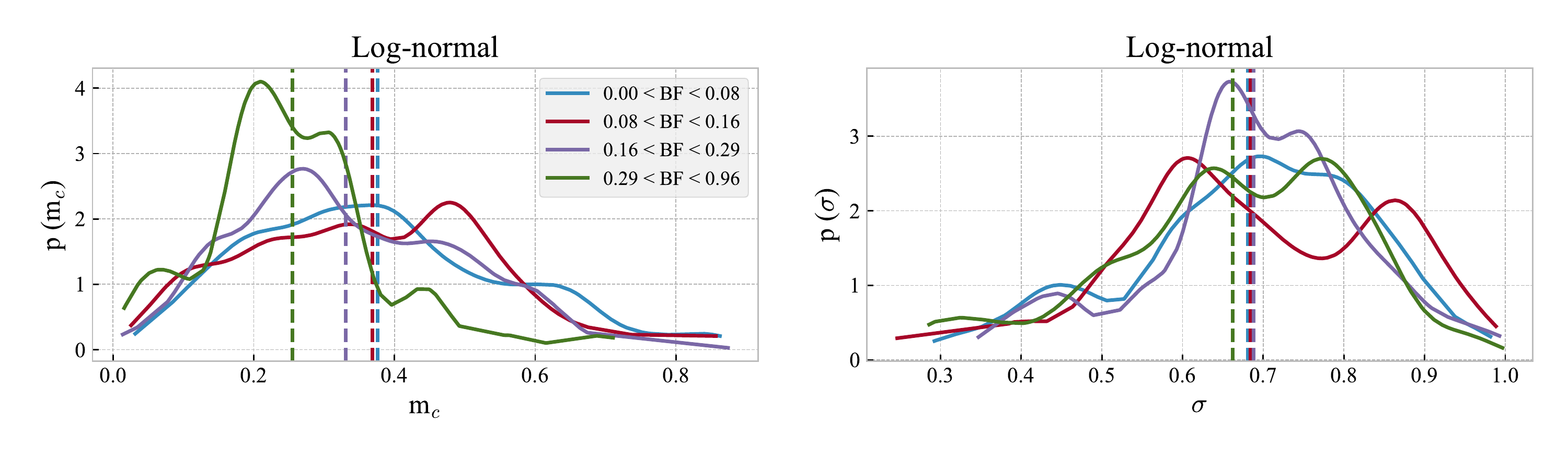}\\
\includegraphics[width=0.99\textwidth]{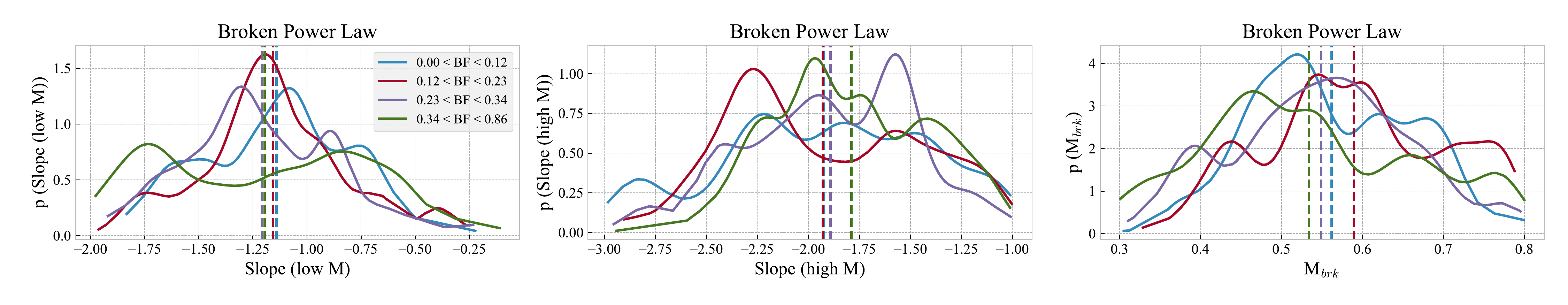}
\caption{Marginal distribution of the fit parameters for different limited to draws corresponding to selected ranges of the binary fraction. Each of the "range slices" in the binary fraction parameter contains the same number of posterior draws. Each slice, indicated by different colors, corresponds to a 25\% quantile. The vertical dashed lines indicate the mean of the corresponding distributions. Top: the SPL case; Center: the LN case; Bottom: the BPL case. \label{fig:binslices}}
\end{figure*}

The two parameters that are observed to correlate the most with the binary fraction are the slope in the SPL model and the characteristic mass, $m_c$ in the LN model. 
The correlation goes in the same direction for both models: a smaller binary fraction corresponds to a more bottom-light IMF. At increasingly high binary fraction, the marginal distributions for the slope in the SPL model and $m_c$ in the LN model become more consistent with what is observed in the Milky Way.
A similar but less prominent correlation is observed for the slopes and the break mass in the BPL case.
Even in this case, when considering a higher binary fraction, the IMF becomes more "bottom-heavy" or more Galaxy-like.

While this is an interesting exercise, we do not have any prior knowledge that allows us to independently constrain the binary fraction, and thus narrow down the uncertainty on the paramters that correlate with it.
Our fitting procedure allows us to marginalize over the uncertain binary fraction and report more realistic uncertainties on the other parameters, than if we were to arbitrarely fix them to, e.g. the typical values observed in the Galaxy of about 30\% \citep{2010ApJS..190....1R}.

A further caveat to consider is the way in which binary pairing is implemented in our simulations. As explained in \cite{2018ApJ...855...20G}, our IMF represents the distribution of system masses. In simulating synthetic CMDs, we draw system masses, $M_{sys}$, from an IMF model, then decide whether a system should be a single star or a binary, according to a probability set by the binary fraction. If the system is labeled as binary, a mass ratio, $q = M_2/M_1$, is drawn from a uniform distribution between 0 and 1. Individual masses are given by $M_1 = \frac{M_{sys}}{1+q}$ and $M_2 = \frac{qM_{sys}}{1+q}$ . This is different than drawing individual masses from the same IMF and pairing randomly. 
The qualitative argument for our choice is that it is unlikely that the two stars in a binary are completely independent draws from the same IMF. For progressively tighter binaries it is reasonable to assume that at formation the two stars affected each other's amount of accreted mass from the protostellar phase.

In determining the mass ratio distribution, we used the fact that for "solar-type" stars, the binary mass ratio is uniformly distributed between 0 and 1 \citep{1991A&A...248..485D,2010ApJS..190....1R}.
Such a distribution may not be valid outside of the Milky Way or in different mass regimes. We thus stress that, while we use a reasonable assumption, the derived fraction of binaries depends on the adopted mass-ratio distribution.

Nevertheless, we performed a sanity check on the adopted rule of binary component pairing.
We verified that our process for generating binaries, which we refer to as correlated draws, does not alter the inferred slope, with respect to a method in which the two stars in a binary are generated from the same single star IMF, referred to as independent draws.
We assume the same input slope for the single star IMF and the system IMF in the independent and correlated case respectively. We then compare the single star (primaries plus secondaries), system, as well as the primary star and secondary star distributions.
Differences appear at the high and low-end of the mass spectrum.
The differences at the low-mass edge are due to the fact that in the correlated case, secondaries can in principle have $M_2 \rightarrow 0$, while a finite limit must be set for extracting individual masses from a power law in the independent case. At the high-mass edge there are differences due to the definitions of upper limits, i.e., we cannot enforce the same limit on both the maximum individual mass and the maximum system mass consistently for both the independent and correlated scenarios.
When limiting the analysis to the mass region accessible to us, i.e., far from both the low-mass and high-mass end of the IMF, no significant differences appear for either the mass ratio distribution, the individual stars distribution, and the system mass distribution between the two methods of generating binaries.
As an example, a 2-samples Kolomogorov-Smirnov test for 1000 draws from each of the independent and correlated cases, gives a p-value of 0.95, thus the null-hypothesis that the distributions of the two samples are the same cannot be rejected.

\section{Summary and conclusions}

%We have for the first time probed a mass regime very close to the hydrogen burning limit for a system external to the Milky Way, the ultra faint dwarf galaxy Coma Berenices. This has allowed us to study its Initial Mass Function down to a mass of $\sim0.23$~M$_{\odot}$. 

We have probed the IMF of the ultra faint dwarf Coma Berenices down to a mass of 0.23 M$_{\odot}$, the lowest mass probed to date in a Milky Way satellite, approaching the hydrogen burning limit.
Our observation have in fact a deeper limit of about 0.17M$_{\odot}$. However the purity of our sample decreases strongly below F160W = 28.6 mag, corresponding to 0.23 M$_{\odot}$. This can be attributed to the limited angular resolution of $HST$ in the near-infrared. We fail to distinguish small, faint background galaxies from Com~Ber stars below such magnitude.

We have explored three possible IMF parameterizations: Single Power Law (SPL), Broken Power Law (BPL) and Log-normal (LN). 
%For all three of these  our findings for the ultra-faint dwarf galaxy Coma Berenices are that the IMF parameter values differs significantly from the Milky Way ones.
%While the results for a single power law model may suggest a more bottom light IMF for Com~Ber than the Milky Way, the discrepancy is reduced when using a broken power law or log-normal model.
%The limited sample size for Com~Ber members hampers a quantitative model comparison between the 3 different parameterizations.% There is however a qualitative suggestion that no turn-over of the IMF is observed at low masses, below 0.5~M$_{\odot}$. This is indicated by both the fact that the single power-law model gives a very good fit to the data and by the fact that for the log-normal and broken power-law models, the characteristic mass values are slightly higher than for the Milky Way. %We cannot exclude a turnover at masses greater than $\sim0.6\, \mathrm{~M}_{\odot}$, but the limited number of stars above such a threshold, as well as the short mass interval above the putative turn-over (due to the turn-off mass being at $\sim0.8 \mathrm{~M}_{\odot}$), make it difficult to demonstrate the existence of such feature.
%We conclude that it is highly unlikely that the Milky Way stars and Coma Berenices ones have been drawn from the same IMF.
Summarizing, we find that:
\begin{itemize}
\item the best-fit results for all three IMF parametrization produce similarly good qualitative agreement with the data,
\item no strong bias due to the adopted mass range is evident in the result for the SPL or LN model, when compared to the results of \cite{2018ApJ...855...20G}
\item the SPL model best fit values for Com~Ber are marginally consistent with the Milky Way value by \cite{1955ApJ...121..161S}. The BPL and LN best fits are fully consistent with the \cite{2001MNRAS.322..231K} and \cite{2003PASP..115..763C} results for the same models for the Milky Way, respectively
\end{itemize}

We discuss that part of the discrepancy between the Milky Way IMF and that for a sample of 6 UFDs, including Com~Ber, claimed in \cite{2018ApJ...855...20G} could be related to using a SPL model, which may lead to artificially shallow slopes, if the underlying IMF is a log-normal and if the log-normal characteristic mass is below the probed mass range \cite[see][]{2017MNRAS.468..319E}.
We note however that extending the mass range with respect to \cite{2018ApJ...855...20G} down to the Milky Way characteristic mass value \citep{2003PASP..115..763C} does not change the slope values significantly.
Background contamination could be an additional factor in making the derived Com~Ber IMF artificially steeper. 
On the other end, 3 UFDs in the \cite{2018ApJ...855...20G} sample of 6 galaxies show a significantly more bottom-light IMF than Com~Ber. This could mean that there is true variance between the UFDs, as quantified by the very different SPL best-fit values, and therefore while some, like Com~Ber, might have an IMF consistent with the Milky Way one, others, like CVn~II, Hercules and Leo~IV, might truly differ from the Galaxy.
Moreover the good fit provided by the SPL model for Com~Ber indicates that there is no evidence of an IMF turnover, at least with the available number of observed stars.

At this stage we are still gathering evidence on whether the IMF in low-metallicity systems, which formed stars 14 Gyr ago when our Universe was a very different environment, is different from the IMF of stars that are currently forming in our own Galaxy.
This leaves the question open for study with future observatories that will become available in the next decade. JWST and WFIRST, above all, will provide both depth to push the probed mass range towards the hydrogen burning limit as well as the field of view to efficiently study nearby, low surface brightness dwarf galaxies; in addition, in the case of JWST, we will be able to exploit the increased angular resolution needed to improve star/galaxy separation at faint limits.

\acknowledgements
Support for program GO-13449 was provided by NASA through a grant from
the Space Telescope Science Institute, which is operated by the
Association of Universities for Research in Astronomy, Inc., under
NASA contract NAS 5-26555.  

\software{DAOPHOT-II package \cite{1987PASP...99..191S}, Victoria-Regina evolutionary code \citep{2014ApJ...794...72V}, MARCS \citep{2008A&A...486..951G} , corner.py code \citep{corner}}

\bibliography{biblio_comberIR}
\bibliographystyle{aasjournal}

\end{document}